\documentclass[10pt,journal,cspaper,compsoc]{IEEEtran}

\usepackage{amsmath,amssymb}
\usepackage{graphicx}
\usepackage{hyperref}
\usepackage{algorithm,algorithmic}
\usepackage{multirow}
\usepackage{cite}
\usepackage{url}
\usepackage{color}
\usepackage{tabularx}
\usepackage{xspace}
\usepackage{booktabs}
\usepackage{listings}

\newcommand{\vct}[1]{\ensuremath{\boldsymbol{#1}}} 

\newcommand{\set}[1]{\ensuremath{\mathcal{#1}}}
\newcommand{\con}[1]{\ensuremath{\mathsf{#1}}}
\newcommand{\T}{\ensuremath{\top}}

\newcommand{\argmax}{\operatornamewithlimits{\arg\,\max}}

\newcommand{\argmin}{\operatornamewithlimits{\arg\,\min}}

\newcommand{\tsum}{\textstyle \sum}

\newcommand{\SVM}{SVM\xspace}
\newcommand{\SSVM}{Sec-SVM\xspace}
\newcommand{\MCS}{MCS-SVM\xspace}
\newcommand{\Drebin}{\textrm{Drebin}\xspace}

\newcommand{\DDrebin}{\textsl{Drebin}\xspace}
\newcommand{\DContagio}{\textsl{Contagio}\xspace}

\newcommand{\manifest}{\texttt{manifest}\xspace}
\newcommand{\dexcode}{\texttt{dexcode}\xspace}

\newcommand{\ie}{{i.e.}\xspace}
\newcommand{\eg}{{e.g.}\xspace}
\newcommand{\etal}{{et al.}\xspace}
\newcommand{\etc}{{etc.}\xspace}

\newcommand{\myparagraph}[1]{\smallskip \noindent \textbf{#1.}}

\begin{document}

\title{Yes, Machine Learning Can Be More Secure! \\ A Case Study on Android Malware Detection}

\author{Ambra~Demontis,~\IEEEmembership{Student Member,~IEEE,}
    Marco~Melis,~\IEEEmembership{Student Member,~IEEE,}
	Battista~Biggio,~\IEEEmembership{Senior Member,~IEEE,}
	Davide~Maiorca,~\IEEEmembership{Member,~IEEE,}
	Daniel~Arp,
	Konrad~Rieck,
        Igino~Corona,
        Giorgio~Giacinto,~\IEEEmembership{Senior Member,~IEEE,}
        and~Fabio~Roli,~\IEEEmembership{Fellow,~IEEE}
\IEEEcompsocitemizethanks{
\IEEEcompsocthanksitem
A.~Demontis (ambra.demontis@diee.unica.it),
M.~Melis (marco.melis@diee.unica.it),
B.~Biggio (battista.biggio@diee.unica.it),
D.~Maiorca (davide.maiorca@diee.unica.it),
I.~Corona (igino.corona@diee.unica.it),
G.~Giacinto (giacinto@diee.unica.it) and
F.~Roli (roli@diee.unica.it) are with the Dept. of Electrical and Electronic Eng., University of Cagliari, Piazza d'Armi, 09123 Cagliari, Italy.
\IEEEcompsocthanksitem D.~Arp (arp@sectubs.de) and K.~Rieck (rieck@sectubs.de) are with the Institute of System Security, Technische Universit{\"a}t Braunschweig, Rebenring 56, 38106 Braunschweig, Germany.}}%

\IEEEcompsoctitleabstractindextext
{\begin{abstract}
To cope with the increasing variability and sophistication of modern attacks, machine learning has been widely adopted as a statistically-sound tool for malware detection.
However, its security against well-crafted attacks has not only been recently questioned, but it has been shown that machine learning exhibits inherent vulnerabilities that can be exploited to evade detection at test time. In other words, machine learning itself can be the weakest link in a security system.
In this paper, we rely upon a previously-proposed attack framework to categorize potential attack scenarios against learning-based malware detection tools, by modeling attackers with different skills and capabilities. We then define and implement a set of corresponding evasion attacks to thoroughly assess the security of Drebin, an Android malware detector.
The main contribution of this work is the proposal of a simple and scalable secure-learning paradigm that mitigates the impact of evasion attacks, while only slightly worsening the detection rate in the absence of attack.
We finally argue that our secure-learning approach can also be readily applied to other malware detection tasks.
\end{abstract}

}

\maketitle

\IEEEdisplaynotcompsoctitleabstractindextext

\IEEEpeerreviewmaketitle

\IEEEraisesectionheading{\section{Introduction}}\label{sect:introduction}

\IEEEPARstart{D}{uring} the last decade, machine learning has been increasingly applied in security-related tasks, in response to the increasing variability and sophistication of modern attacks~\cite{BarKayOorSom10,PenGatSarMol12,AafDuYin13,rieck14-drebin,LinNeuPla15}.
One relevant feature of machine-learning approaches is their ability to \emph{generalize}, \ie, to potentially detect never-before-seen attacks, or variants of known ones.
However, as first pointed out by Barreno \etal~\cite{barreno06-asiaccs,barreno10}, machine-learning algorithms have been designed under the assumption that training and test data follow the same underlying probability distribution, which makes them vulnerable to well-crafted attacks violating this assumption. This means that machine learning itself can be the weakest link in the security chain~\cite{arce03}. Subsequent work has confirmed this intuition, showing that machine-learning techniques can be significantly affected by carefully-crafted attacks exploiting knowledge of the learning algorithm; \eg, skilled attackers can manipulate data at test time to \emph{evade} detection, or inject \emph{poisoning} samples into the training data to mislead the learning algorithm and subsequently cause misclassification errors~\cite{NewKarSon06,PerDagLeeFogSha06,VenBluSon08,huang11,biggio12-icml,biggio15-icml,biggio13-ecml,wang14-usenix,srndic14}.

In this paper, instead, we show that one can leverage machine learning to improve system security, by following an \emph{adversary-aware} approach in which the machine-learning algorithm is designed from the \emph{ground up} to be more resistant against evasion.
We further show that designing adversary-aware learning algorithms according to this principle, as advocated in~\cite{biggio14-tkde,biggio14-ijprai}, does not
necessarily require one to trade classification accuracy in the absence of carefully-crafted attacks for improving security.

\begin{figure*}[t]
\begin{center}
\includegraphics[width=0.75\textwidth]{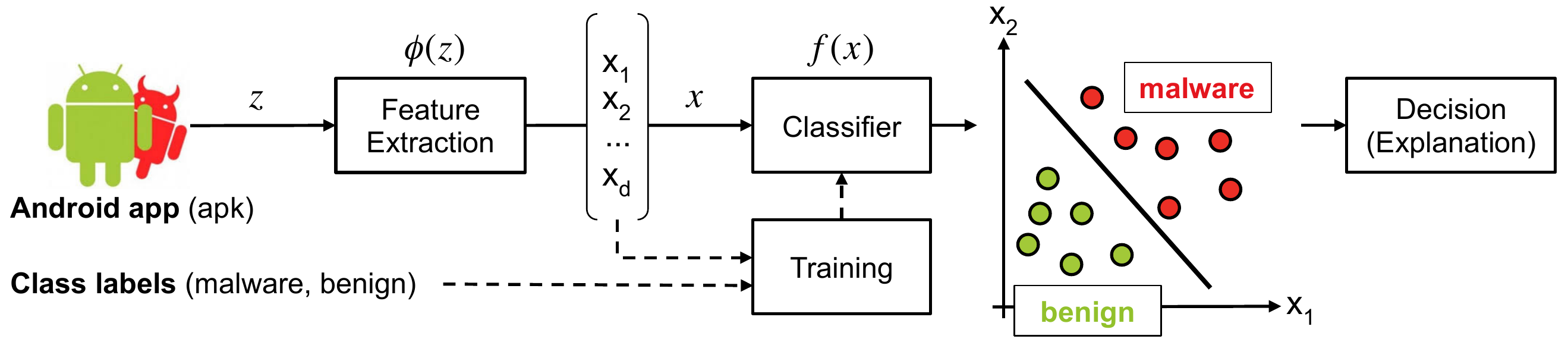}
\vspace{-8pt}
\caption{A schematic representation of the architecture of \Drebin. First, applications are represented as vector in a $\con d$-dimensional feature space. A linear classifier is then trained on an available set of labeled application, to discriminate between malware and benign applications. During classification, unseen applications are evaluated by the classifier. If its output $f(\vct x) \geq 0$, they are classified as malware, and as benign otherwise. \Drebin also provides an interpretation of its decision, by highlighting the most suspicious (or benign) features that contributed to the decision~\cite{rieck14-drebin}.}
\vspace{-16pt}
\label{fig:system-arch}
\end{center}
\end{figure*}

We consider Android malware detection as a case study for our approach.
The relevance of this task is witnessed by the fact that Android
has become the most popular mobile operating system,
with more than a billion users around the world, while the number of malicious applications targeting them has also grown simultaneously: anti-virus vendors detect thousands of new malware samples daily,
and there is still no end in sight~\cite{ZhoJia12,LinNeuWeiFraVeePla14}.
Here we focus our analysis on \Drebin (Sect.~\ref{sect:drebin}), \ie, a machine-learning approach that relies on \emph{static analysis} for an efficient detection of Android malware directly on the mobile device~\cite{rieck14-drebin}.

Notably, in this work we do not consider attacks that can completely defeat static analysis~\cite{MosKruKir07}, like those based on packer-based encryption~\cite{Yang2015} and advanced code obfuscation~\cite{collberg97,Rastogi2015,hoffmann16-codaspy,hoffmann16-tr,reaves16}.
The main reason is that such techniques may leave detectable traces, suggesting the use of a more appropriate system for classification; \eg, the presence of system routines that perform dynamic loading of libraries or classes, potentially hiding embedded malware, demands for the use of dynamic analysis for a more reliable classification. 
For this reason, in this paper we aim to improve the security of \Drebin against \emph{stealthier} attacks, 
\ie, carefully-crafted malware samples that evade detection without exhibiting significant evidence of manipulation. 

To perform a well-crafted security analysis of \Drebin and, more generally, of Android malware detection tools against such attacks, we exploit an adversarial framework (Sect.~\ref{sect:threat-model}) based on previous work on \emph{adversarial machine learning}~\cite{barreno06-asiaccs,barreno10,huang11,biggio14-tkde,biggio14-ijprai}. We focus on the definition of different classes of \emph{evasion} attacks, corresponding to \emph{attack scenarios} in which the attacker exhibits an increasing capability of manipulating the input data, and level of knowledge about the targeted system.
To simulate evasion attacks in which the attacker does not exploit any knowledge of the targeted system,
we consider some obfuscation techniques that are not specifically targeted against \Drebin, by running an analysis similar to that reported in~\cite{maiorca15-cose}. 
To this end, we make use of the commercial obfuscation tool \texttt{DexGuard},\footnote{\url{https://www.guardsquare.com/dexguard}} which has been originally designed to make reverse-engineering of benign applications more difficult. The obfuscation techniques exploited by this tool are discussed in detail in Sect.~\ref{sect:obfuscation_techniques}. 
Note that, even if considering obfuscation attacks is out of the scope of this work, \texttt{DexGuard} only partially obfuscates the content of Android applications. For this reason, the goal of this analysis is simply to empirically assess whether the static analysis performed by \Drebin remains effective when Android applications are not thoroughly obfuscated, or when obfuscation is not targeted.

The main contribution of this work is the proposal of an \emph{adversary-aware} machine-learning detector against \emph{evasion attacks} (Sect.~\ref{sect:sec-learning}), inspired from the proactive design approach advocated in the area of adversarial machine learning~\cite{biggio14-tkde,biggio14-ijprai}.
The secure machine-learning algorithm proposed in this paper is completely novel. With respect to previous techniques for \emph{secure learning}~\cite{bruckner12,globerson06-icml,kolcz09,biggio10-ijmlc}, it is able to retain computational efficiency and scalability on large datasets (as it exploits a linear classification function), while also being well-motivated from a more theoretical perspective.
We empirically evaluate our method on real-world data (Sect.~\ref{sect:exp}), including an adversarial security evaluation based on the simulation of the proposed evasion attacks. We show that our method outperforms state-of-the-art classification algorithms, including \emph{secure} ones, without losing significant accuracy in the absence of well-crafted attacks, and can even guarantee some degree of robustness against \texttt{DexGuard}-based obfuscations. We finally discuss the main limitations of our work (Sect.~\ref{sect:limitations}), and future research challenges, including how to apply the proposed approach to other malware detection tasks (Sect.~\ref{sect:conclusion}).

\vspace{-3pt}
\section{Android Malware Detection}
\label{sect:drebin}

In this section, we give some background on Android applications.
We then discuss \Drebin and its main limitations.

\vspace{-2pt}
\subsection{Android Background}
\label{sect:android_malware}

Android is the most used mobile operating system. Android applications are in the \texttt{apk} format, \ie, a zipped archive containing two files: the Android \manifest and classes.dex. Additional \texttt{xml} and resource files are respectively used to define the application layout, and to provide additional functionality or multimedia content. As \Drebin only analyzes the Android \manifest and classes.dex files, below we provide a brief description of their characteristics.

\myparagraph{Android Manifest} The \manifest file holds information about the application structure. Such structure is organized in \emph{application components}, \ie, parts of code that perform specific actions; \eg, one component might be associated to a screen visualized by the user (\emph{activity}) or to the execution of audio in the background (\emph{services}). The actions of each component are further specified through \emph{filtered intents}; \eg, when a component sends data to other applications, or is invoked by a browser. Special types of components are \emph{entry points}, \ie, activities, services and receivers that are loaded when requested by a specific filtered intent (\eg, an activity is loaded when an application is launched, and a service is activated when the device is turned on).
The \manifest also contains the list of \emph{hardware components} and \emph{permissions} requested by the application to work (\eg, Internet access).

\myparagraph{Dalvik Bytecode (\dexcode)} The classes.dex file
contains the compiled source code of an application. It contains all the user-implemented methods and classes.  
Classes.dex might contain specific API
calls that can access sensitive resources such as personal
contacts (\emph{suspicious calls}). Moreover, it contains all system-related,
\emph{restricted API calls} whose functionality require
\emph{permissions} (\eg, using the Internet). Finally, this
file can contain references to \emph{network addresses} that
might be contacted by the application.

\vspace{-2pt}
\subsection{Drebin}

\Drebin conducts multiple steps and can be executed directly on the
mobile device, as it performs a lightweight static
analysis of  Android applications. The extracted
features are used to embed applications into a high-dimensional
vector space and train a classifier on a set of labeled data. An
overview of the system architecture is given in
Fig.~\ref{fig:system-arch}. In the following, we describe the single
steps in more detail.

\subsubsection{Feature Extraction}
    Initially, \Drebin performs a static analysis of a set of available
    Android applications,\footnote{We use here a modified
    version of \Drebin that performs a static
    analysis based on the \texttt{Androguard} tool, available at:\\ \url{https://github.com/androguard/androguard}.} to construct a suitable feature space.
    All features extracted by \Drebin are presented as \emph{strings} and
    organized in 8 different feature sets, as listed in Table \ref{tab:feature_sets}.
    \begin{table}[t]
          \caption{Overview of feature sets.} \vspace{-5pt}
      \centering
    \begin{tabular}{ l|ll }
      \toprule
      \multicolumn{3}{ c }{\textbf{Feature sets}} \\
      \midrule
      \multirow{4}{*}{\texttt{manifest}}
        & $S_{1}$ & Hardware components \\
        & $S_{2}$ & Requested permissions \\
        & $S_{3}$ & Application components \\
        & $S_{4}$ & Filtered intents \\
      \midrule
      \multirow{4}{*}{\texttt{dexcode}}
        & $S_{5}$ & Restricted API calls \\
        & $S_{6}$ & Used permission \\
        & $S_{7}$ & Suspicious API calls \\
        & $S_{8}$ & Network addresses \\
      \bottomrule
    \end{tabular}\vspace{-10pt}
    \label{tab:feature_sets}
    \end{table}
      Android applications are then mapped onto the feature space as follows.
      Let us assume that an Android application (\ie, an \texttt{apk} file) is
      represented as an object $\vct z \in \set Z$, being $\set Z$
      the abstract space of all \texttt{apk} files.
      We then denote with $\Phi : \set Z \mapsto \set X$ a function
      that maps an \texttt{apk} file $\vct z$ to a $\con d$-dimensional feature vector $\vct
      x = ( x^{1}, \ldots, x^{\con d} )^{\T} \in \set X=\{0,1\}^{\con d}$,
      where each feature is set to 1 (0) if the corresponding \emph{string} is present (absent) in
      the \texttt{apk} file $\vct z$.
      An application encoded in feature space may thus look like the following:
      \[
          \vct x = \Phi( \vct z) \mapsto
         \begin{pmatrix}
          \cdots \\ 0\\ 1\\
          \cdots \\ 1\\ 0\\
          \cdots\\
         \end{pmatrix}
         \begin{array}{ll}
         \cdots & \multirow{4}{*}{\hspace{-1mm}\bigg \} $S_2$ }\\
         \texttt{\small permission::SEND\_SMS} \\
         \texttt{\small permission::READ\_SMS}\\
         \cdots & \multirow{4}{*}{\hspace{-1mm}\bigg \} $S_5$ }\\
         \texttt{\small api\_call::getDeviceId}\\
         \texttt{\small api\_call::getSubscriberId}\\
         \cdots & \\
         \end{array}
      \]

   \subsubsection{Learning and Classification} \label{sect:learning-classification}

      Once Android applications are represented as feature vectors,
      \Drebin learns a \emph{linear} Support Vector Machine (SVM) classifier~\cite{vapnik95,vapnik95-book}
      to discriminate between the class of benign and malicious samples.
      Linear classifiers are generally expressed in terms of
      a linear function $f : \set X \mapsto \mathbb R$, given as:
       \begin{equation}
        f(\vct x) = \vct w^{\T}\vct x + b \, ,
      \label{eq:standard-svm}
      \end{equation}
      where  $\vct w \in \mathbb R^{\con d}$ denotes the vector of \emph{feature weights},
      and $b \in \mathbb R$ is the so-called \emph{bias}.
      These parameters, to be optimized during training, identify a hyperplane in feature space, which separates the two classes.
            During classification, unseen applications are then classified as malware if $f(\vct x) \geq 0$, and as benign otherwise.

      During training, we are given a set of labeled samples $\set D = \{(\vct
      x_{i}, y_{i})\}_{i=1}^{\con n}$, where $\vct x_{i}$ denotes an
      application in feature space, and $y_{i} \in \{-1,+1\}$ its label,
      being $-1$ and $+1$ the benign and malware class, respectively.
      The SVM learning algorithm is then used to find the parameters
      $\vct w,b$ of Eq.~\eqref{eq:standard-svm}, by solving the following optimization problem:
      \begin{equation}
      \min_{\vct w, b} \set L (\set D, f) = \underbrace{\tfrac{1}{2} \vct w^{\T} \vct w}_{R(f)} + C \underbrace{\tsum_{i=1}^{\con n} \max(0, 1-y_{i}f(\vct x_{i}))}_{L(f,\set D)} \, ,
      \label{eq:svm-learning}
      \end{equation}
      where $L(f, \set D)$ denotes a loss function computed on the training data (exhibiting higher values if samples in $\set D$ are not correctly classified by $f$), $R(f)$ is a regularization term to avoid overfitting (\ie, to avoid that the classifier overspecializes its decisions on the training data, losing generalization capability on unseen data), and $C$ is a trade-off parameter.
      As shown by the above problem, the SVM exploits an $\ell_{2}$ regularizer on the feature weights and the so-called \emph{hinge loss} as the loss function. This allows the SVM algorithm to learn a hyperplane that separates the two classes with the highest \emph{margin}~\cite{vapnik95,vapnik95-book}.
      Note that the above formulation is quite general, as it represents different learning algorithms, depending on the chosen regularizer and loss function~\cite{cherkassky09}.

\subsection{Limitations and Open Issues}

Although \Drebin has shown to be capable of detecting malware with high accuracy,
it exhibits intrinsic vulnerabilities that might be exploited by an attacker to evade detection.
Since \Drebin has been designed to run directly on the mobile device,
its most obvious limitation is the lack of a dynamic analysis.
Unfortunately, static analysis has clear limitations, as it is not
possible to analyze malicious code that is downloaded or decrypted
at runtime, or code that is thoroughly obfuscated~\cite{MosKruKir07,collberg97,Yang2015,Rastogi2015,hoffmann16-codaspy,hoffmann16-tr,reaves16}.
For this reason, considering such attacks would be irrelevant for the scope of our work.
Our focus is rather to understand and to improve the security properties of learning algorithms
against specifically-targeted attacks, in which the amount of manipulations performed by the attacker is limited.
The rationale is that the manipulated malware samples should not only evade detection, but it should also be difficult to detect traces of their adversarial manipulation.
Although these limitations have been also discussed in~\cite{rieck14-drebin}, the effect of carefully-targeted attacks against \Drebin has never been studied before. For this reason, in the following, we introduce an attack framework to provide a systematization of different, potential evasion attacks under limited adversarial manipulations. Then, we present a systematic evaluation of these attacks on \Drebin, and a novel learning algorithm to alleviate their effects in practice.

\section{Attack Model and Scenarios}
\label{sect:threat-model}

To perform a thorough security assessment of learning-based malware detection systems, we rely upon an attack model originally defined
in~\cite{biggio14-tkde,biggio14-ijprai}. It is grounded on the popular taxonomy of Barreno \etal~\cite{barreno06-asiaccs,barreno10,huang11}, which categorizes potential attacks against machine-learning algorithms along three axes: \emph{security violation},  \emph{attack specificity} and  \emph{attack influence}. The attack model exploits this taxonomy to define a number of potential attack scenarios that may be incurred by the system
during operation, in terms of explicit assumptions on the attacker's goal, knowledge of the system, and capability of manipulating the input data.

\vspace{-2pt}
\subsection{Attacker's Goal} It is defined in terms of the desired security violation and the so-called attack specificity.

\myparagraph{Security violation} Security can be compromised by violating system \emph{integrity}, if malware samples are undetected;
system \emph{availability}, if benign samples are misclassified as malware; or \emph{privacy}, if the system leaks confidential information about its users.

\myparagraph{Attack specificity} It can be \emph{targeted} or \emph{indiscriminate}, depending on whether the attacker is interested in having some specific samples misclassified (\eg, a specific malware sample to infect a particular device), or if any misclassified sample meets her goal (\eg, if the goal is to launch an indiscriminate attack campaign).

\vspace{-2pt}
\subsection{Attacker's Knowledge} The attacker may have different levels of knowledge of the targeted system~\cite{biggio14-tkde,biggio14-ijprai,barreno06-asiaccs,barreno10,huang11,srndic14}. In particular, she may know completely, partially, or do not have any information at all about:
($i$) the training data $\set D$;
($ii$) the feature extraction/selection algorithm $\Phi$, and the corresponding feature set $\set X$, \ie, how features are computed from data, and selected;
($iii$) the learning algorithm $\set L(\set D, f)$, along with the decision function $f(\vct x)$ (Eq.~\ref{eq:standard-svm}) and, potentially, even its (trained) parameters $\vct w$ and $b$. In some applications, the attacker may also exploit feedback on the classifier's decisions to improve her knowledge of the system, and, more generally, her attack strategy~\cite{biggio14-tkde,biggio14-ijprai,barreno06-asiaccs,huang11}.

\vspace{-2pt}
\subsection{Attacker's Capability} \label{subsect:capability} It consists of defining the \emph{attack influence} and how the attacker can manipulate data.

\myparagraph{Attack Influence} It can be \emph{exploratory}, if the attacker only manipulates  data at test time, or \emph{causative}, if she can also contaminate the training data (\eg, this may happen if a system is periodically retrained on data collected during operation that can be modified by an attacker)~\cite{barreno06-asiaccs,huang11,biggio14-tkde}.

\myparagraph{Data Manipulation} It defines how samples (and features) can be modified, according to application-specific constraints; \eg, which feature values can be incremented or decremented without compromising the exploitation code embedded in the \emph{apk} file. In many cases, these constraints can be encoded in terms of distances in feature space, computed between the source malware data and its manipulated versions~\cite{dalvi04,lowd05,globerson06-icml,teo08,bruckner12,biggio13-ecml}.
We refer the reader to Sect.~\ref{sect:malware-manipulation} for a discussion on how  \Drebin features can be modified.

\vspace{-2pt}
\subsection{Attack Strategy} The attack strategy defines how the attacker implements her activities, based on the hypothesized goal, knowledge, and capabilities.
To this end, we characterize the attacker's knowledge in terms of a space $\Theta$ that encodes knowledge of the data $\set D$, the feature space $\set X$, and the classification function $f$. Accordingly, we can represent the scenario in which the attacker has perfect knowledge of the attacked system as a vector $\vct \theta=(\set D,\set X, f) \in \Theta$.
We characterize the attacker's capability by assuming that an initial set of samples $\set A$ is given, and that it is modified according to a space of possible modifications $\Omega(\set A)$.
Given the attacker's knowledge $\vct \theta\in\Theta$ and a set of manipulated attacks $\set A^{\prime} \in \Omega(\set A) \subseteq \set Z$, the attacker's goal can be characterized in terms of an objective function $\set W (\set A^{\prime}, \vct \theta) \in \mathbb R$ which evaluates the extent to which the manipulated attacks $\set A^{\prime}$ meet the attacker's goal.
The optimal attack strategy can be thus given as:
\begin{eqnarray}
\set A^{\star} =\textstyle \argmax_{\set A'\in\Omega(\set A)} \, \set W(\set A'; \vct \theta) \, .
\label{eq:opt-attack}
\end{eqnarray}
Under this formulation, one can characterize different attack scenarios. The two main ones often considered in adversarial machine learning are referred to as classifier {\bf evasion} and {\bf poisoning}~\cite{biggio13-ecml,biggio12-icml,biggio15-icml,barreno06-asiaccs,barreno10,huang11,biggio14-tkde,biggio14-ijprai}.
In the remainder of this work we focus on \emph{classifier evasion}, while we refer the reader to~\cite{biggio14-tkde,biggio15-icml} for further details on \emph{classifier poisoning}.

\subsection{Evasion Attacks} \label{sect:evasion-attacks}

In an evasion attack, the attacker manipulates malicious samples at test time to have them misclassified as benign by a trained classifier, without having \emph{influence} over the training data.
The attacker's goal thus amounts to violating system \emph{integrity}, either with a \emph{targeted} or with an \emph{indiscriminate} attack, depending on whether the attacker is targeting a specific machine or running an indiscriminate attack campaign.
More formally, evasion attacks can be written as:
\begin{eqnarray}
\vct z^{\star} = \argmin_{\vct z^{\prime} \in \Omega( \vct z )} \hat f(\Phi(\vct z^{\prime})) = \argmin_{\vct z^{\prime} \in \Omega( \vct z )} \hat{\vct w}^{\T}\vct x^{\prime} \, ,
\label{eq:evasion}
\end{eqnarray}
where $\vct x^{\prime} = \Phi(\vct z^{\prime})$ is the feature vector associated to the modified attack sample $\vct z^{\prime}$, and $\hat{\vct w}$ is the weight vector estimated by the attacker (\eg, from the surrogate classifier $\hat f$).
With respect to Eq.~\eqref{eq:opt-attack}, one can consider here one sample at a time, as they can be independently modified.

The above equation essentially tells the attacker which features should be modified to maximally decrease the value of the classification function, \ie, to maximize the probability of evading detection~\cite{biggio14-tkde,biggio13-ecml}. Note that, depending on the manipulation constraints $\Omega( \vct z )$ (\eg, if the feature values are bounded), the set of features to be manipulated is generally different for each malicious sample.

In the following, we consider different evasion scenarios, according to the framework discussed in the previous sections. In particular, we discuss five distinct attack scenarios, sorted for increasing level of attacker's knowledge. Note that, when the attacker knows more details of the targeted system, her estimate of the classification function becomes more reliable, thus facilitating the evasion task (in the sense of requiring less manipulations to the malware samples). 

\subsubsection{Zero-effort Attacks}
 This is the standard scenario in which malware data is neither obfuscated nor modified at all. From the viewpoint of the attacker's knowledge, this scenario is characterized by an empty knowledge-parameter vector $\vct \theta = ()$.

\subsubsection{DexGuard-based Obfuscation Attacks} \label{sect:obf}
As another attack scenario in which the attacker does not exploit any knowledge of the attacked system, for which $\vct \theta = ()$, we consider a setting similar to that reported in~\cite{maiorca15-cose}.
In particular, we assume that the attacker attempts to evade detection by performing invasive code transformations on the \texttt{classes.dex} file, using the commercial Android obfuscation tool \texttt{DexGuard}. Note that this tool is designed to ensure protection against disassembling/decompiling attempts in benign applications, and not to obfuscate the presence of malicious code; thus, despite the introduction of many changes in the executable code, it is not clear whether and to what extent the obfuscations implemented by this tool may be effective against a learning-based malware detector like \Drebin, \ie, how they will affect the corresponding feature values and classification output.
The obfuscations implemented by \texttt{DexGuard} are described more in detail in Sect.~\ref{sect:obfuscation_techniques}.

\subsubsection{Mimicry Attacks}
Under this scenario, the attacker is assumed to be able to collect a surrogate dataset including malware and benign samples, and to know the feature space. Accordingly, $\vct \theta = (\hat{\set D}, \set X)$. In this case, the attack strategy amounts to manipulating malware samples to make them as close as possible to the benign data (in terms of conditional probability distributions or, alternatively, distance in feature space).
To this end, in the case of \Drebin (which uses binary feature values), we can assume that the attacker still aims to minimize Eq.~\eqref{eq:evasion}, but estimates each component of $\hat{\vct w}$ independently for each feature as $\hat w_{k} = p(\hat x_{k} = 1 | y=+1) - p(\hat x_{k}=1 | y=-1)$, $k=1,\ldots,\con d$. This will indeed induce the attacker to add (remove) first features which are more frequently present (absent) in benign files, making the probability distribution of malware samples closer to that of the benign data.
It is worth finally remarking that this is a more sophisticated mimicry attack than those commonly used in practice, in which an attacker is usually assumed to merge a malware application with a benign one~\cite{srndic14,ZhoJia12}.

\subsubsection{Limited-Knowledge (LK) Attacks}
In addition to the previous case, here the attacker knows the learning algorithm $\set L$ used by the targeted system, and can learn a surrogate classifier on the available data. The knowledge-parameter vector can be thus encoded as $\vct \theta = (\hat{\set D}, \set X, \hat f)$, being $\hat f$ the surrogate classifier used to approximate the true $f$. In this case, the attacker exploits the estimate of $\hat{\vct w}$ obtained from the surrogate classifier $\hat f$ to construct the evasion samples, according to Eq.~\eqref{eq:evasion}.

\subsubsection{Perfect-Knowledge (PK) Attacks}
This is the worst-case setting in which also the targeted classifier is known to the attacker, \ie, $\vct \theta = (\set D, \set X, f)$. Although it is not very likely to happen in practice that the attacker gets to know even the trained classifier's parameters (\ie, $\vct w$ and $b$ in Eq.~\ref{eq:standard-svm}), this setting is particularly interesting as it provides an upper bound on the performance degradation incurred by the system under attack, and can be used as reference to evaluate the effectiveness of the system under the other simulated attack scenarios.

\subsection{Malware Data Manipulation} \label{sect:malware-manipulation}

As stated in Sect.~\ref{subsect:capability}, one has to discuss how the attacker can manipulate malware applications to create the corresponding evasion attack samples.
To this end, we consider two main settings in our evaluation, detailed below.

\myparagraph{Feature Addition}  Within this setting, the attacker can
independently inject (\ie, set to 1) every feature.

\myparagraph{Feature Addition and Removal} This scenario simulates a more powerful attacker that can inject every feature, and also remove (\ie, set to 0) features from the \dexcode.

\smallskip
These settings are motivated by the fact that malware has to be manipulated
to evade detection, but its semantics and intrusive functionality must be preserved.
In this respect, \emph{feature addition} is generally a safe operation,
in particular, when injecting \manifest features
    (\eg, adding
    permissions does not influence any existing application
    functionality).
    With respect to the \dexcode, one may also safely
    introduce information that is not actively executed, by adding code
      after \texttt{return}
    instructions (\emph{dead code}) or with methods that are never called by
    any \texttt{invoke} type instructions.
	Listing~\ref{lst:smali-example} shows an
    example where a URL feature is introduced by adding a method that
    is never invoked in the code.

\lstset{
  frame = lrbt,
  numbers = left,
  aboveskip=3mm,
  belowskip=3mm,
  showstringspaces=false,
  columns=flexible,
  basicstyle={\small\ttfamily},
  numbers=none,
  numberstyle=\tiny\color{gray},
  keywordstyle=\color{blue},
  commentstyle=\color{dkgreen},
  stringstyle=\color{mauve},
  breaklines=true,
  breakatwhitespace=true,
  tabsize=3
}

\begin{lstlisting}[caption=Smali code to add a URL feature.,
label={lst:smali-example}, captionpos=b]
.method public addUrlFeature()V
  .locals 2
  const-string v1, "http://www.example.com"
  invoke-direct {v0, v1},
  Ljava/net/URL;-><init>(Ljava/lang/String;)V
  return-void
.end method
\end{lstlisting}
\vspace{-5pt}

However, this only applies when such information is not directly executed by the application, and could be stopped at the parsing level by analyzing only the methods belonging to the application \emph{call graph}.
In this case, the attacker would be enforced to change the executed code, and this requires considering additional and stricter constraints. For example, if she wants to add a suspicious API call to a \dexcode method that is executed by the application, she should adopt virtual machine registers that have not been used before by the application. Moreover, the attacker should pay attention to possible artifacts or undesired functionalities that are brought by the injected calls, which may influence the semantics of the original program. Accordingly, injecting a large number of features may not always be feasible.

\emph{Feature removal} is even a more complicated operation. Removing permissions from the \manifest is not possible, as this would limit the application functionality. The same holds for intent filters. Some application component names can be changed but, as stated in Sect.~\ref{sect:obfuscation_techniques}, this operation is not easy to be automatically performed: the attacker must ensure that the application component names in the \dexcode are changed accordingly, and must not modify any of the entry points. Furthermore, the feasible changes may only slightly affect the whole \manifest structure (as shown in our experiments with automated obfuscation tools).
With respect to the \dexcode, multiple ways can be exploited to remove its features; \eg,
it is possible to hide IP addresses (if
    they are stored as strings) by encrypting them with the
    introduction of additional functions, and decrypting them at
    runtime. Of course, this should be done by avoiding the addition of
    features that are already used by the system (\eg,
    function calls that are present in the training data). 
    
    With respect to suspicious and restricted API calls, the attacker should encrypt the method or the class invoking them. However, this could introduce other calls that might increase the suspiciousness of the application. Moreover, one mistake at removing such API references might completely destroy the application functionality. The reason is that Android uses a \emph{verification} system to check the integrity of an application during execution (\eg, it will close the application, if a register passed as a parameter to an API call contains a wrong type), and chances of compromising this behavior increase if features are deleted carelessly.


For the aforementioned reasons, performing a fine-grained evasion attack that changes a lot of features may be very difficult in practice, without compromising the malicious application functionality. In addition, another problem for the attacker is getting to know precisely which features should be added or removed, which makes the construction of evasion attack samples even more complicated.

\section{DexGuard-based Obfuscation Attacks}
\label{sect:obfuscation_techniques}

Although commercial obfuscators are designed to protect benign applications against reverse-engineering attempts, it has been recently shown that they can also be used to evade anti-malware detection systems~\cite{maiorca15-cose}.  
We thus use \texttt{DexGuard}, a popular obfuscator for Android, to simulate attacks in which no specific knowledge of the targeted system is exploited, as discussed in Sect.~\ref{sect:obf}. 
Recall that, although considering obfuscation attacks is out of the scope of this work, the obfuscation techniques implemented by \texttt{DexGuard} do not completely obfuscate the code. For this reason, we aim to understand whether this may make static analysis totally ineffective, and how it affects our strategy to improve classifier security.
A brief description of the \texttt{DexGuard}-based obfuscation attacks is given below.


\myparagraph{Trivial obfuscation}  This strategy changes the names of \emph{implemented} application packages, classes, methods and fields, by replacing them with random
characters. Trivial obfuscation also 
performs negligible modifications to some \manifest
features by renaming some application components that are \emph{not}
entry points (see Sect.~\ref{sect:android_malware}). As the application functionality must be preserved, Trivial obfuscation does not rename any system API or method imported from native libraries. Given that \Drebin mainly extracts information from system APIs, we expect that its detection capability will be only barely affected by this obfuscation. 

\myparagraph{String Encryption} This strategy encrypts strings defined in
the \dexcode with the instruction \texttt{const-string}. Such strings
can be visualized during the application execution, or may be used as
variables. Thus, even if they are retrieved through an
identifier, their value must be preserved during the program execution. For
this reason, an additional method is added to decrypt them at runtime, 
when required. 
This obfuscation tends to
remove URL features (S8) that are stored as strings in the \dexcode.
Features corresponding to the decryption routines extracted by \Drebin (S7) are instead not affected, as the decryption routines added by \texttt{DexGuard} do not belong to the system APIs.

\myparagraph{Reflection} This obfuscation technique uses the Java Reflection API to replace
\texttt{invoke-type} instructions with calls that belong to the \texttt{Java.lang.Reflect} class. 
The main effect of this action is destroying the application call graph. However, this technique does not affect the system API names, as they do not get encrypted during the process. It is thus reasonable to expect that most of the features extracted by \Drebin will remain unaffected.

\myparagraph{Class Encryption} This is the most invasive obfuscation
strategy, as it encrypts all the application classes, except entry-point ones
(as they are required to load the application externally).
The encrypted classes are decrypted at runtime
by routines that are added during the obfuscation phase. 
Worth noting, the class encryption performed by \texttt{DexGuard} does not completely encrypt the application. For example, classes belonging to the API components contained in the \manifest are not encrypted, as this would most likely compromise the application functionality. For the same reason, the \manifest itself is preserved. Accordingly, it still possible to extract static features using \Drebin, and analyze the application.
Although out of the scope of our work, it is still worth remarking here that using packers (\eg, ~\cite{Yang2015}) to perform full dynamic loading of the application classes might completely evade static analysis. 

\myparagraph{Combined Obfuscations} The aforementioned strategies
can also be combined to produce additional obfuscation techniques.
As in~\cite{maiorca15-cose}, we will consider three additional techniques in our experiments,
by respectively combining ($i$) trivial and string encryption,
($ii$) adding reflection to them, and ($iii$) adding class encryption to the former three.

\vspace{-5pt}
\section{Adversarial Detection}
\label{sect:sec-learning}

In this section, we introduce an adversary-aware approach to improve the robustness of \Drebin against carefully-crafted data manipulation attacks.
As for \Drebin, we aim to develop a simple, lightweight and scalable approach. For this reason, the use of non-linear classification functions with computationally-demanding learning procedures is not suitable for our application setting. We have thus decided to design a linear classification algorithm with improved security properties, as detailed in the following.

\subsection{Securing Linear Classification}

As in previous work~\cite{kolcz09,biggio10-ijmlc}, we aim to improve the security of our linear classification system by enforcing learning of more \emph{evenly-distributed} feature weights, as this would intuitively require the attacker to manipulate more features to evade detection.
Recall that, as discussed in Sect.~\ref{sect:malware-manipulation}, if a large number of features has to be manipulated to evade detection, it may not even be possible to construct the corresponding malware sample without compromising its malicious functionality.
With respect to the work in~\cite{kolcz09,biggio10-ijmlc}, where different heuristic implementations were proposed to improve the so-called \emph{evenness} of feature weights (see Sect.~\ref{sect:exp}), we propose here a more principled approach, derived from the idea of \emph{bounding} classifier sensitivity to feature changes.

We start by defining a measure of \emph{classifier sensitivity} as:
\begin{equation}
\Delta f(\vct x,\vct x^{\prime}) = \frac{f(\vct x)-f(\vct x^{\prime})}{ \| \vct x-\vct x^{\prime} \|} = \frac{\vct w^{\T} (\vct x-\vct x^{\prime})}{ \| \vct x-\vct x^{\prime} \|}  \, ,
\end{equation}
which evaluates the decrease of $f$ when a malicious sample $\vct x$ is manipulated as $\vct x^{\prime}$, with respect to the required amount of modifications, given by $\| \vct x-\vct x^{\prime} \|$.

Let us assume now, without loss of generality, that $\vct w$ has unary $\ell_{1}$-norm and that features are normalized in $[0,1]$.\footnote{Note that this is always possible without affecting system performance, by dividing $f$ by $\| \vct w \|_{1}$, and normalizing feature values on a compact domain before classifier training.}
We also assume that, for simplicity, the $\ell_{1}$-norm is used to evaluate $\| \vct x-\vct x^{\prime} \|$.
Under these assumptions, it is not difficult to see that $\Delta f \in \left[ \tfrac{1}{\con d}, 1 \right]$, where the minimum is attained for equal absolute weight values (regardless of the amount of modifications made to $\vct x$), and the maximum is attained when only one weight is not null, confirming the intuition that more \emph{evenly-distributed} feature weights should improve classifier security under attack.
This can also be shown by selecting $\vct x,\vct x^{\prime}$ to maximize $\Delta f(\vct x, \vct x^{\prime})$:
\begin{align}
\Delta f(\vct x,\vct x^{\prime})  \leq \tfrac{1}{\con K} \textstyle{\sum_{k=1}^{\con K}} |w_{(k)}| \leq \max_{j=1,\ldots,\con d} |w_{j}| = {\| \vct w \|_{\infty}} .
\end{align}
Here, $\con K = \| \vct x - \vct x^{\prime}\|$ corresponds to the number of modified features and $|w_{(1)}|, \ldots, |w_{(\con d)}|$ denote the weights sorted in descending order of their absolute values, such that we have $|w_{(1)}| \geq \ldots \geq |w_{(\con d)}|$.
The last inequality shows that, to minimize classifier sensitivity to feature changes, one can minimize the $\ell_{\infty}$-norm of $\vct w$.
This in turn tends to promote solutions which exhibit the same absolute weight values (a well-known effect of $\ell_{\infty}$ regularization~\cite{Boyd-Vandenberghe-Convex-2004}).

This is a very interesting result which has never been pointed out in the field of adversarial machine learning. We have shown that regularizing our learning algorithm by penalizing the $\ell_{\infty}$ norm of the feature weights $\vct w$ can improve the \emph{security} of linear classifiers, yielding classifiers with more \emph{evenly-distributed} feature weights. This has only been intuitively motivated in previous work, and implemented with heuristic approaches~\cite{kolcz09,biggio10-ijmlc}.
As we will show in Sect.~\ref{sect:exp}, being derived from a more principled approach,
our method is not only capable of finding more \emph{evenly-distributed} feature weights with respect to the heuristic approaches in~\cite{kolcz09,biggio10-ijmlc}, but it is also able to outperform them in terms of security.

It is also worth noting that our approach preserves \emph{convexity} of the objective function minimized by the learning algorithm. This gives us the possibility of deriving computationally-efficient training algorithms with (potentially strong) convergence guarantees.
As an alternative to considering an additional term to the learner's objective function $\set L$,
one can still control the $\ell_{\infty}$-norm of $\vct w$ by adding a box constraint on it. This is a well-known property of convex optimization~\cite{Boyd-Vandenberghe-Convex-2004}.
As we may need to apply different upper and lower bounds to different feature sets, depending on how their values can be manipulated, we prefer to follow the latter approach.

\subsection{Secure SVM Learning Algorithm}
\label{sect:secsvm-alg}

According to the previous discussion, we define our Secure SVM learning algorithm (\SSVM) as:
\begin{eqnarray}
\label{eq:obj-ssvm}
\min_{\vct w, b} &&  \tfrac{1}{2}\vct {w}^{\T} \vct w + C \tsum_{i=1}^{\con n} \max \left( 0, 1-y_{i}f(\vct x_{i}) \right) \, , \\
\label{eq:constr-ssvm}
{\rm s. t.} &&  w^{\rm lb}_{k} \leq w_{k} \leq w^{\rm ub}_{k} \, , \, k = 1, \ldots, \con d \, .
\end{eqnarray}
Note that this optimization problem is identical to Problem~\eqref{eq:svm-learning}, except for the presence of a box constraint on $\vct w$.
The lower and upper bounds on $\vct w$ are defined by the vectors $\vct w^{\rm lb} = (w^{\rm lb}_{1}, \ldots, w^{\rm lb}_{\con d})$ and $\vct w^{\rm ub} = (w^{\rm ub}_{1}, \ldots, w^{\rm ub}_{\con d})$, which should be selected with a suitable procedure (see Sect.~\ref{sect:sec-par-tuning}).
For notational convenience, in the sequel we will also denote the constraint given by Eq.~\eqref{eq:constr-ssvm} compactly as $\vct w \in \set W \subseteq \mathbb R^{\con d}$.

The corresponding learning algorithm is given as Algorithm~\ref{alg:secure-classifier}.
It is a constrained variant of Stochastic Gradient Descent (SGD) that also considers a simple line-search procedure to tune the gradient step size during the optimization.
SGD is a lightweight gradient-based algorithm for efficient learning on very large-scale datasets, based on approximating the subgradients of the objective function using a single sample or a small subset of the training data, randomly chosen at each iteration~\cite{zhang04-sgd,bottou10-sgd}.
In our case, the subgradients of the objective function (Eq.~\ref{eq:obj-ssvm}) are given as:
\begin{eqnarray}
\label{eq:grad1} \nabla_{\vct w} \set L &\approxeq& \vct w + C \, \tsum_{i \in \set S} \nabla_{\ell}^{i} \,  \vct x_{i} \, ,\\
\label{eq:grad2} \nabla_{b} \set L &\approxeq& C \, \tsum_{i \in \set S} \nabla_{\ell}^{i} \, ,
\end{eqnarray}
where $\set S$ denotes the subset of the training samples used to compute the approximation, and
$\nabla_{\ell}^{i}$ is the gradient of the hinge loss with respect to $f(\vct x_{i})$, which equals $-y_{i}$, if $y_{i}f(\vct x_{i}) < 1$, and $0$ otherwise.

One crucial issue to ensure quick convergence of SGD is the choice of the initial gradient step size $\eta^{(0)}$, and of a proper \emph{decaying function} $s(t)$, \ie, a function used to gradually reduce the gradient step size during the optimization process. As suggested in~\cite{zhang04-sgd,bottou10-sgd}, these parameters should be chosen based on preliminary experiments on a subset of the  training data. Common choices for the function $s(t)$ include linear and exponential decaying functions.

We conclude this section by pointing out that our formulation is quite general; one may indeed select different combinations of loss and regularization functions to train different, secure  variants of other linear classification algorithm. Our \SSVM learning algorithm is only an instance that considers the hinge loss and $\ell_{2}$ regularization, as the standard SVM~\cite{vapnik95,vapnik95-book}.
It is also worth remarking that, as the lower and upper bounds become smaller in absolute value, our method tends to yield (\emph{dense}) solutions with weights equal to the upper or to the lower bound. A similar effect is obtained when minimizing the $\ell_{\infty}$ norm directly~\cite{Boyd-Vandenberghe-Convex-2004}.

We conclude from this analysis that there is an implicit trade-off between \emph{security} and \emph{sparsity}: while a sparse learning model ensures an efficient description of the learned decision function, it may be easily circumvented by just manipulating a few features. By contrast, a secure learning model relies on the presence of many, possibly redundant, features that make it harder to evade the decision function, yet at the price of a dense representation.

\begin{algorithm}[tb]
  \caption{\SSVM Learning Algorithm}
  \label{alg:secure-classifier}
  \textbf{Input:} $\set D = \{ \vct x_{i}, y_{i}\}_{i=1}^{\con n}$, the training data;
  	$C$, the regularization parameter;
	$\vct w^{\rm lb},\vct w^{\rm ub}$, the lower and upper bounds on $\vct w$;
	$| \set S |$, the size of the sample subset used to approximate the subgradients;
	$\eta^{(0)}$, the initial gradient step size;
	$s(t)$, a decaying function of $t$;
	and $\varepsilon > 0$, a small constant.\\
  \textbf{Output:} $\vct w, b$, the trained classifier's parameters.

    \begin{algorithmic}[1]
    \STATE{Set iteration count $t \gets 0$.}
    \STATE{Randomly initialize $\vct v^{(t)} = (\vct w^{(t)}, b^{(t)}) \in \set W \times \mathbb R$.}
    \STATE{Compute the objective function $\set L(\vct v^{(t)})$ using Eq.~\eqref{eq:obj-ssvm}.}
    \REPEAT
     \STATE{Compute $\left (\nabla_{\vct w} \set L, \nabla_{b} \set L \right)$ using Eqs.~\eqref{eq:grad1}-\eqref{eq:grad2}.}
      \STATE{Increase the iteration count $t \gets t+1$.}
      \STATE{Set $\eta^{(t)} \gets  \gamma \, \eta^{(0)}  s(t)$, performing a line search on $\gamma$.}
      \STATE{Set $\vct w^{(t)} \gets \vct w^{(t-1)} - \eta^{(t)} \nabla_{w} \set L$.}
      \STATE{Project $\vct w^{(t)}$ onto the feasible (box) domain $\set W$.}
      \STATE{Set $b^{(t)} \gets b^{(t-1)} - \eta^{(t)} \nabla_{b} \set L$.}
      \STATE{Set $\vct v^{(t)} = (\vct w^{(t)}, b^{(t)})$.}
      \STATE{Compute the objective function $\set L(\vct v^{(t)})$ using Eq.~\eqref{eq:obj-ssvm}.}
     \UNTIL{$ | \set L (\vct v^{(t)}) - \set L(\vct v^{(t-1)})  | < \varepsilon$}
      \STATE{\textbf{return:} $\vct w = \vct w^{(t)}$, and $b = b^{(t)}$.}
  \end{algorithmic}

\end{algorithm}

\subsection{Parameter Selection} \label{sect:sec-par-tuning}

To tune the parameters of our classifiers, as suggested in~\cite{biggio14-tkde,zhang16-tcyb}, one should not only optimize accuracy on a set of collected data, using traditional performance evaluation techniques like cross validation or bootstrapping. More properly, one should optimize a trade-off between accuracy and \emph{security}, by accounting for the presence of potential, unseen attacks during the validation procedure.
Here we optimize this trade-off, denoted with $r(f_{\vct \mu}, \set D)$, as:
\begin{equation}
\vct \mu^{\star} = \textstyle \argmax_{\vct \mu} \, r(f_{\vct \mu}, \set D) = A(f_{\vct \mu}, \set D) + \lambda S(f_{\vct \mu}, \set D) \, ,
\label{eq:secxval_tradeoff}
\end{equation}
where we denote
with $f_{\vct \mu}$ the classifier learned with parameters $\vct \mu$ (\eg, for our \SSVM, $\vct \mu = \{ C, \vct w^{\rm lb}, \vct w^{\rm ub} \}$),
with $A$ a measure of classification accuracy in the absence of attack (estimated on $\set D$),
with $S$ an estimate of the classifier security under attack (estimated by simulating attacks on $\set D$), and with $\lambda$ a given trade-off parameter.

Classifier security can be evaluated by considering distinct attack settings, or a different amount of modifications to the attack samples.
In our experiments, we will optimize security in a worst-case scenario, \ie, by simulating a PK evasion attack with both feature addition and removal. We will then average the performance under attack over an increasing number of modified features $m \in [1, \con M]$. More specifically, we will measure security as:
\begin{equation}
S = \tfrac{1}{\con M} \tsum_{m=1}^{\con M} A(f_{\vct \mu}, \set D^{\prime}_{k}) \, ,
\label{eq:secxval_security}
\end{equation}
where $\set D^{\prime}_{k}$ is obtained by modifying a maximum of $m$ features in each malicious sample in the \emph{validation set},\footnote{Note that, as in standard performance evaluation techniques, data is split into distinct training-validation pairs, and then performance is averaged on the distinct validation sets. As we are considering \emph{evasion attacks}, training data is not affected during the attack simulation, and only malicious samples in the validation set are thus modified.} as suggested by the PK evasion attack strategy.

\section{Experimental Analysis} \label{sect:exp}

In this section, we report an experimental evaluation of our proposed secure learning algorithm (\SSVM) by testing it under different evasion scenarios (see Sect.~\ref{sect:evasion-attacks}).

\myparagraph{Classifiers} We compare our \SSVM approach with the standard \Drebin implementation (denoted with \SVM), and with a previously-proposed technique that improves security of linear classifiers by using a Multiple Classifier System (MCS) architecture to obtain a linear classifier with more evenly-distributed feature weights~\cite{kolcz09,biggio10-ijmlc}.
To this end, multiple linear classifiers are learned by sampling uniformly from the training set (a technique known as \emph{bagging}~\cite{breiman96-bagging}) and by randomly subsampling the feature set, as suggested by the \emph{random subspace method}~\cite{ho98}.
The classifiers are then combined by averaging their outputs, which is equivalent to using a linear classifier whose weights and bias are the average of the weights and biases of the base classifiers, respectively.
With this simple trick, the computational complexity at test time remains thus equal to that of a single linear classifier~\cite{biggio10-ijmlc}.
As we use linear SVMs as the base classifiers, we denote this approach with \MCS.
We finally consider a version of our \SSVM trained using only \manifest features that we call \SSVM (M). The reason is to verify whether considering only features, which can not removed, limits closely mimicking benign data and thereby yields a more secure system.

\myparagraph{Datasets} In our experiments, we use two distinct datasets.
The first (referred to as \DDrebin) includes the data used in~\cite{rieck14-drebin}, and consists of $121,329$ benign applications and $5,615$ malicious samples, labeled using the VirusTotal service. A sample is labeled as malicious if it is detected by at least five anti-virus scanners, whereas it is labeled as benign if no scanner flagged it as malware.
The second (referred to as \DContagio) includes the data used in~\cite{maiorca15-cose}, and consists of about $1,500$ malware samples, obtained from the \textsl{MalGenome}\footnote{\url{http://www.malgenomeproject.org/}} and the \textsl{Contagio Mobile Minidump}\footnote{\url{http://contagiominidump.blogspot.com/}} datasets.
Such samples have been obfuscated with the seven obfuscation techniques described in Sect.~\ref{sect:obfuscation_techniques}, yielding a total of about $10,500$ samples.

\myparagraph{Training-test splits} We average our results on 10 independent runs. In each repetition, we randomly select 60,000 applications from the \DDrebin dataset, and split them into two equal sets of 30,000 samples each, respectively used as the training set and the surrogate set (as required by the LK and mimicry attacks discussed in Sect.~\ref{sect:evasion-attacks}).
As for the test set, we use all the remaining samples from \DDrebin.
In some attack settings (detailed below), we replace the malware data from \DDrebin in each test set with the malware samples from \DContagio.
This enables us to evaluate the extent to which a classifier (trained on some data) preserves its performance in detecting malware from \emph{different} sources.\footnote{Note however that a number of malware samples in \DContagio are also included in the \DDrebin dataset.}

\myparagraph{Feature selection} When running \Drebin on the given datasets, more than one million of features are found. For computational efficiency, we retain the most discriminant $\con d^{\prime}$ features, for which $|p( x_{k} = 1 | y=+1) - p( x_{k}=1 | y=-1)|,~k = 1,\ldots,\con d$, exhibits the highest values (estimated on training data).
In our case, using only $\con d^{\prime} = 10,000$ features does not significantly affect the accuracy of \Drebin. This is consistent with the recent findings in~\cite{RoyDeloLiHer15}, as it is shown that only a very small fraction of features is significantly discriminant, and usually assigned a non-zero weight by \Drebin (\ie, by the SVM learning algorithm).
For the same reason, the sets of selected features turned out to be the same in each run. Their sizes are reported in Table~\ref{tab:drebin-freqs}.

\begin{table}[t]
  \normalsize
    \caption{Number of features in each set for SVM, \SSVM, and \MCS. Feature set sizes for the \SSVM (M) using only \manifest features are reported in brackets. For all classifiers, the total number of selected features is $\con d^{\prime} = 10,000$.}
  \centering
\begin{tabular}{ |c| cr |c| cr }
  \toprule
  \multicolumn{6}{ c }{\textbf{Feature set sizes}} \\
  \midrule
  \parbox[t]{2mm}{\multirow{4}{*}{\rotatebox[origin=c]{90}{\texttt{manifest}}}} &
    $S_{1}$ & 13 (21) &
    \parbox[t]{2mm}{\multirow{4}{*}{\rotatebox[origin=c]{90}{\texttt{dexcode}}}} &
    $S_{5}$ & 147 (0) \\
    &$S_{2}$ & 152  (243)  &   &$S_{6}$ & 37 (0) \\
    &$S_{3}$ & 2,542 (8,904) &   &$S_{7}$ & 3,029 (0) \\
    &$S_{4}$ & 303 (832)  &  &$S_{8}$ & 3,777 (0)\\
  \bottomrule
\end{tabular}
\label{tab:drebin-freqs}
\end{table}

\myparagraph{Parameter setting} We run some preliminary experiments on a subset of the training set and noted that changing $C$ did not have a significant impact on classification accuracy for all the SVM-based classifiers (except for higher values, which cause overfitting). Thus, also for the sake of a fair comparison among different SVM-based learners, we set $C=1$ for all classifiers and repetitions.
For the \MCS classifier, we train $50$ base linear SVMs on random subsets of $80\%$ of the training samples and $50\%$ of the features, as this ensures a sufficient diversification of the base classifiers,  providing more evenly-distributed feature weights.
The bounds of the \SSVM are selected through a 5-fold cross-validation, following the procedure explained in Sect.~\ref{sect:sec-par-tuning}.
In particular, we set each element of $\vct w^{\text{ub}}$ ($\vct w^{\text{lb}}$) as $w^{\text{ub}}$ ($w^{\text{lb}}$), and optimize the two scalar values $(w^{\text{ub}}, w^{\text{lb}}) \in \{0.1, 0.5, 1\} \times \{-1, -0.5, -0.1\}$.
As for the performance measure $A(f_{\vct \mu}, \set D)$ (Eq.~\ref{eq:secxval_tradeoff}), we consider the Detection Rate (DR) at 1\% False Positive Rate (FPR), while the security measure $S(f_{\vct \mu}, \set D)$ is simply given by Eq.~\eqref{eq:secxval_security}.
We set $\lambda = 10^{-2}$ in Eq.~\eqref{eq:secxval_tradeoff} to avoid worsening the detection of both benign and malware samples in the absence of attack to an unnecessary extent. Finally, as explained in Sect.~\ref{sect:secsvm-alg}, the parameters of Algorithm~\ref{alg:secure-classifier} are set by running it on a subset of the training data, to ensure quick convergence, as $\eta^{(0)} = 0.5$, $\gamma \in \{ 10, 20, \ldots, 70\}$ and $s(t) = 2^{-0.01t} / \sqrt \con n$.

\myparagraph{Evasion attack algorithm} We discuss here the algorithm used to implement our advanced evasion attacks. For linear classifiers with binary features, the solution to Problem~\eqref{eq:evasion} can be found as follows. 
First, the estimated weights $\hat{\vct w}$ have to be sorted in descending order of their absolute values, along with the feature values $\vct x$ of the initial malicious sample. 
This means that, if the sorted weights and features are denoted respectively with $\hat w_{(1)}, \ldots, \hat w_{(\con d)}$ and $x_{(1)}, \ldots, x_{(\con d)}$, then $| \hat w_{(1)} | \geq \ldots \geq | \hat w_{(\con d)} |$.
Then, for $k = 1, \ldots, \con d$:
\begin{itemize}
 \item \emph{if} $x_{(k)} = 1$ and $\hat w_{(k)} > 0$ (and the feature is not in the \texttt{manifest} sets S1-S4), \emph{then} $x_{(k)}$ is set to zero;
 \item \emph{if} $x_{(k)} = 0$ and $\hat w_{(k)} < 0$, \emph{then} $x_{(k)}$ is set to one;
 \item \emph{else} $x_{(k)}$ is left unmodified.
 \end{itemize}
If the maximum number of modified features has been reached, the for loop is clearly stopped in advance.

%

\subsection{Experimental Results}

We present our results by reporting the performance of the given classifiers against ($i$) zero-effort attacks, ($ii$) obfuscation attacks, and ($iii$) advanced evasion attacks, including PK, LK and mimicry attacks, with both feature addition, and feature addition and removal (see Sects.~\ref{sect:evasion-attacks}-\ref{sect:malware-manipulation}).

\begin{figure}[t]
\centering
\includegraphics[width=0.168\textheight]{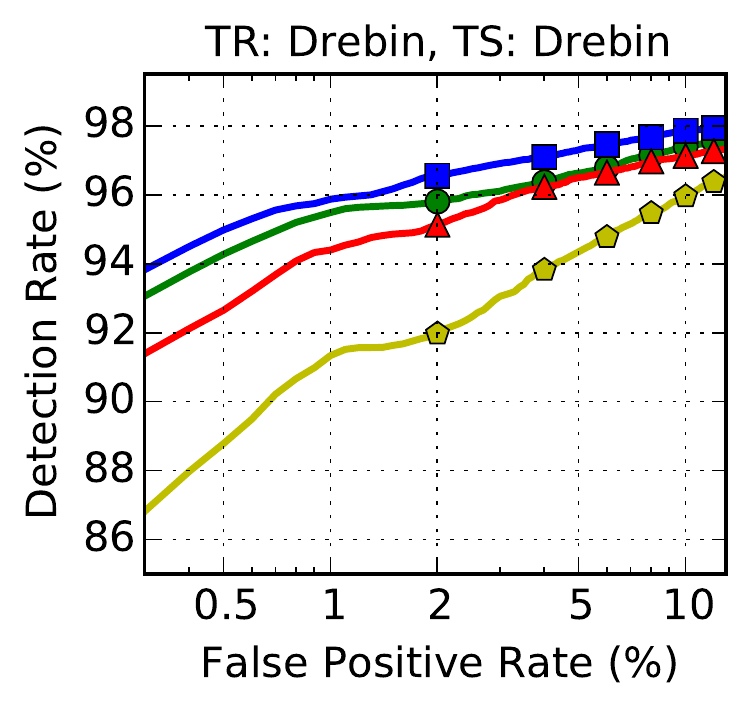}
\includegraphics[width=0.171\textheight]{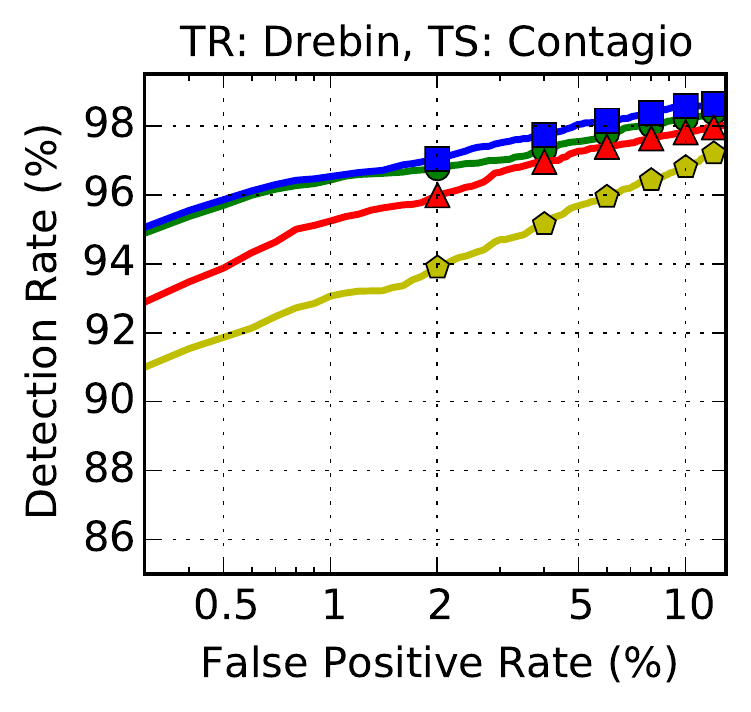}
\includegraphics[width=0.45\textwidth]{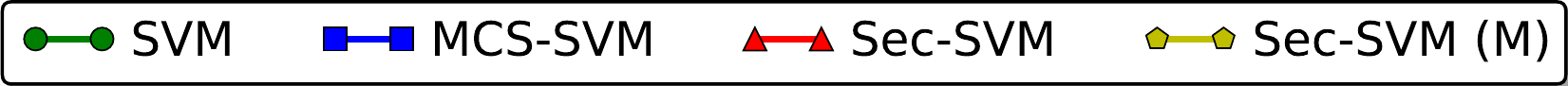}
\vspace{-5pt}
\caption{Mean ROC curves on \DDrebin (\emph{left}) and \DContagio (\emph{right}) data, for classifiers trained on \DDrebin data.}\vspace{-6pt}
\label{fig:roc-drebin-noevasion}
\end{figure}

\begin{figure}[t]
\centering
\includegraphics[width=0.43\textwidth]{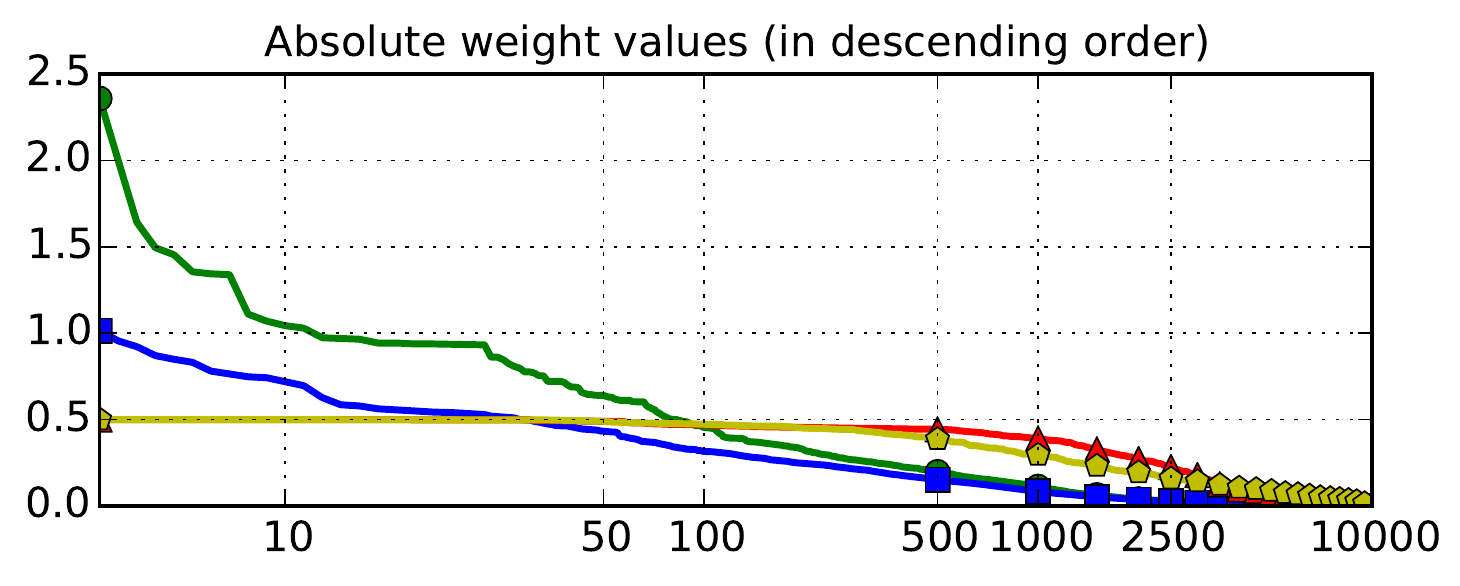}
\includegraphics[width=0.45\textwidth]{figs/legend-horizontal.pdf}
\vspace{-5pt}
\caption{Absolute weight values in descending order (\ie, $|w_{(1)}| \geq \ldots \geq |w_{(\con d)}|$), for each classifier (averaged on 10 runs). Flatter curves correspond to more evenly-distributed weights, \ie, more secure classifiers.}\vspace{-6pt}
\label{fig:wdist-drebin}
\end{figure}

\myparagraph{Zero-effort attacks}  Results for the given classifiers in the absence of attack are reported in the ROC curves of Fig.~\ref{fig:roc-drebin-noevasion}. They report the Detection Rate (DR, \ie, the fraction of correctly-classified malware samples) as a function of the False Positive Rate (FPR, \ie, the fraction of misclassified benign samples) for each classifier.
We consider two different cases: ($i$) using both training and test samples from \DDrebin (\emph{left} plot); and ($ii$) training on \DDrebin and testing on \DContagio  (\emph{right} plot), as previously discussed.
Notably, \MCS achieves the highest DR (higher than $96\%$ at $1\%$ FPR) in both settings, followed by \SVM and \SSVM, which only slightly worsen the DR. \SSVM (M) performs instead significantly worse. In Fig.~\ref{fig:wdist-drebin}, we also report the absolute weight values (sorted in descending order) of each classifier, to show that \SSVM classifiers yield more evenly-distributed weights, also with respect to \MCS.

\begin{figure*}[t]
\centering
\includegraphics[width=0.22\textwidth]{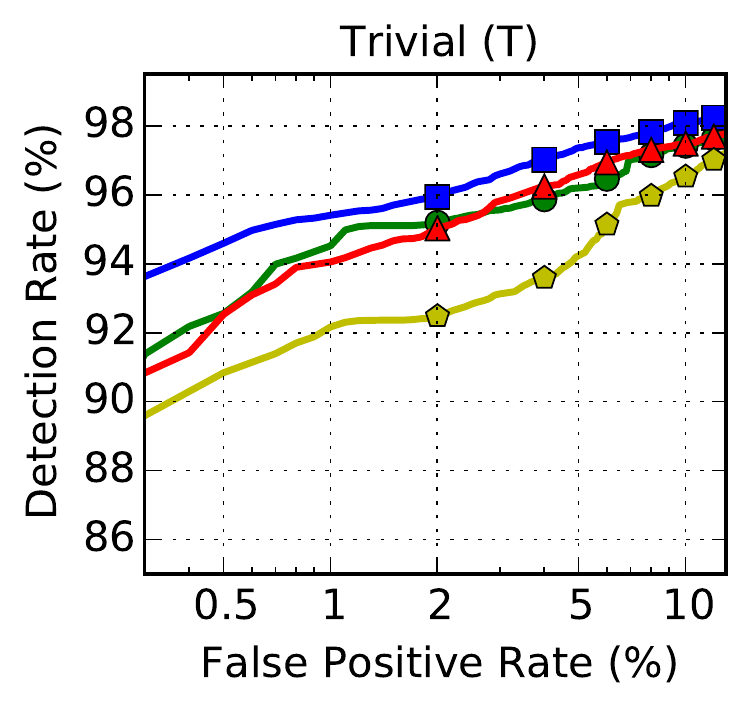}
\includegraphics[width=0.22\textwidth]{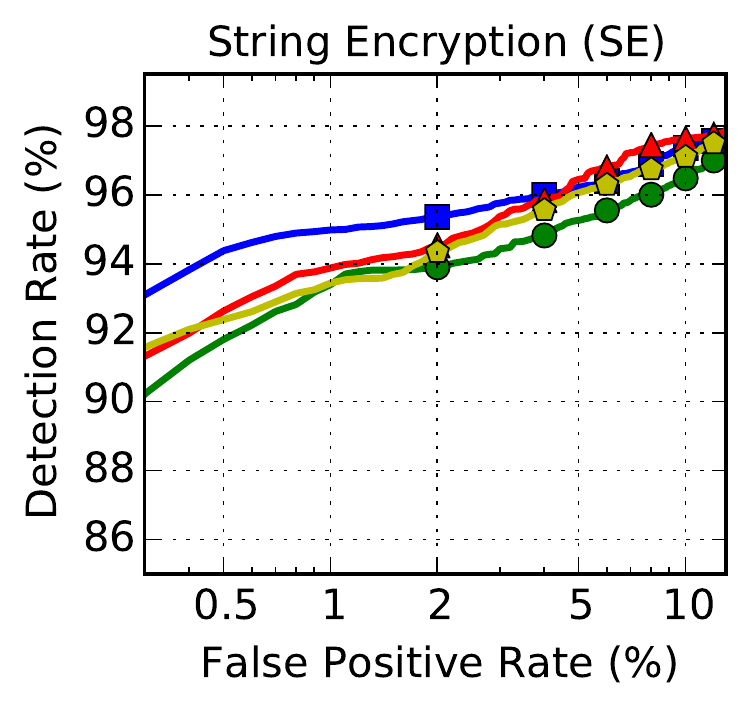}
\includegraphics[width=0.22\textwidth]{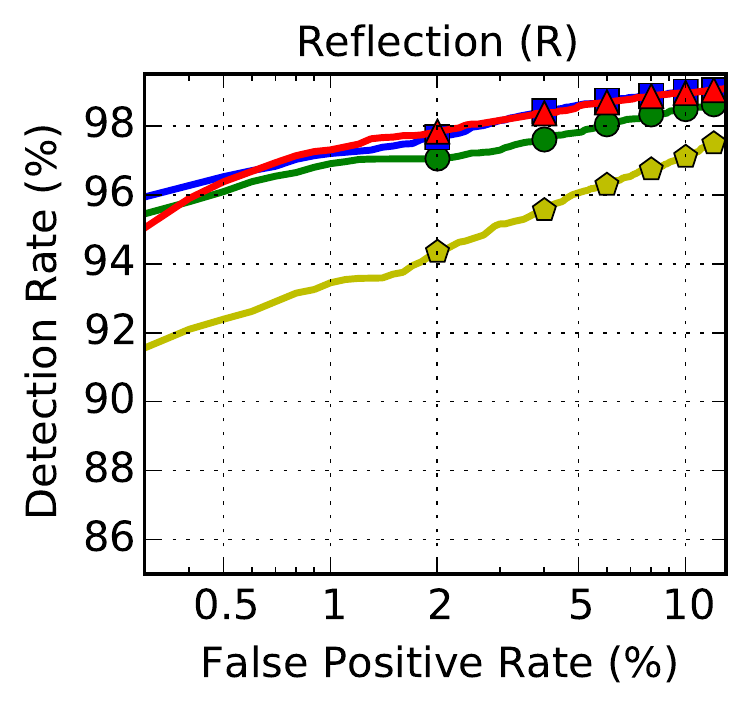}
\includegraphics[width=0.22\textwidth]{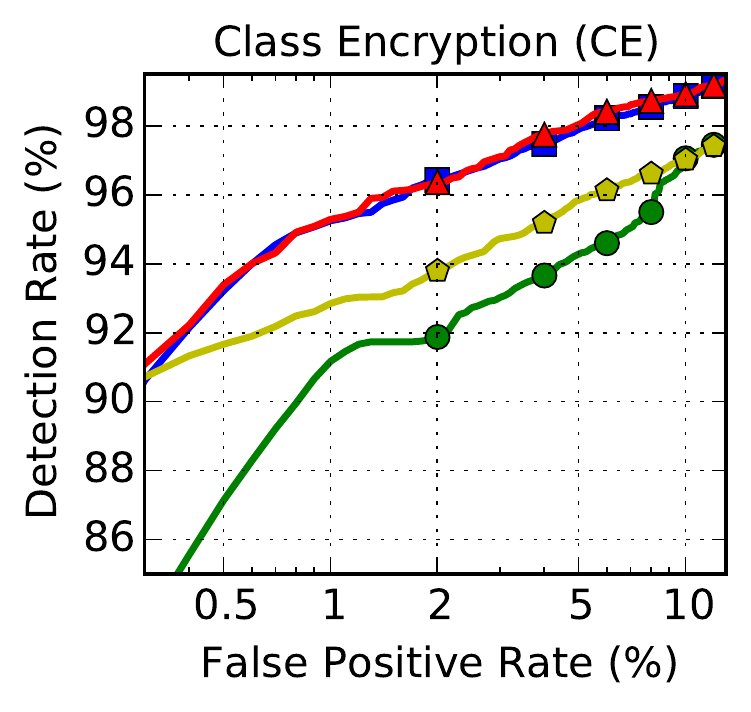}
\hspace{0.015\textwidth}
\raisebox{15.5mm}{\includegraphics[width=0.13\textwidth]{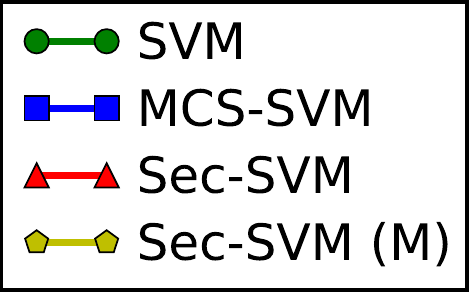} }
\hspace{0.01\textwidth}
\includegraphics[width=0.22\textwidth]{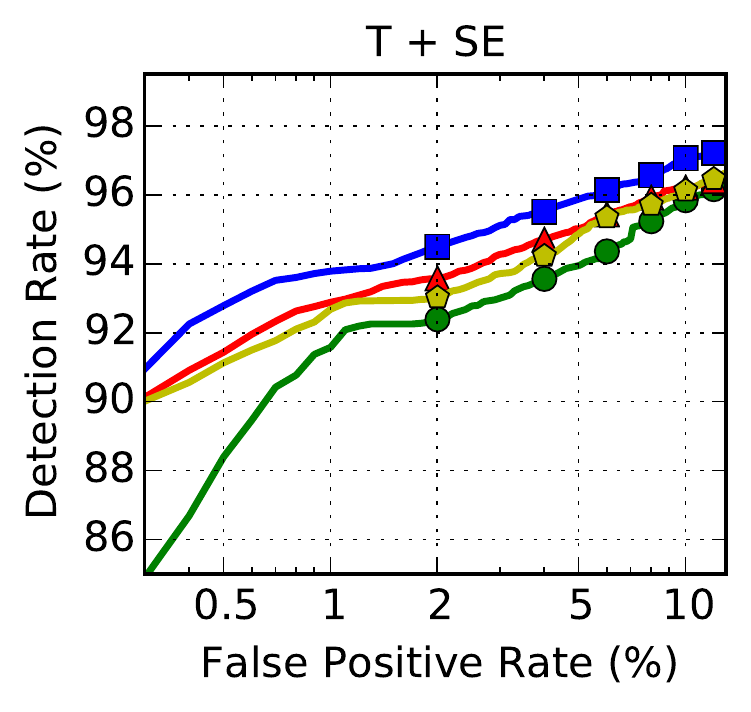}
\includegraphics[width=0.22\textwidth]{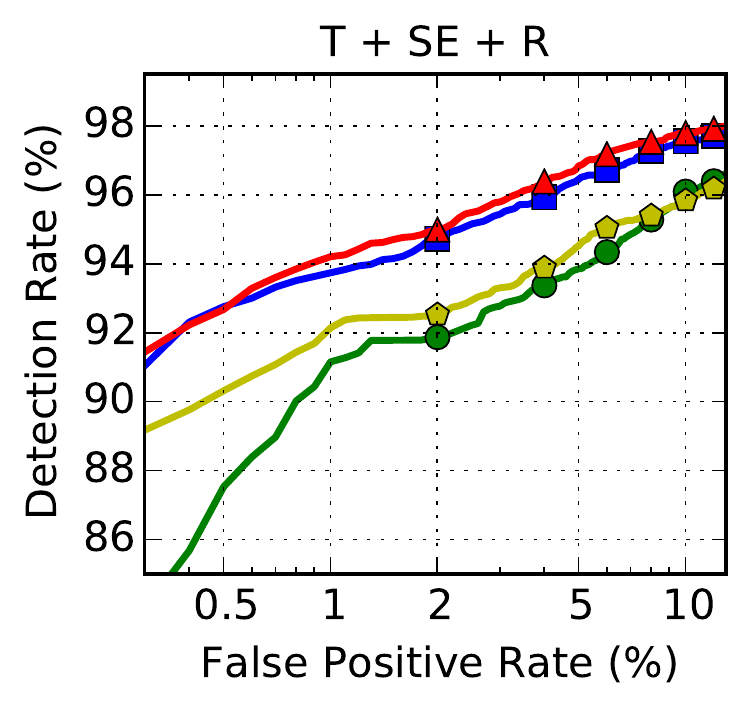}
\includegraphics[width=0.22\textwidth]{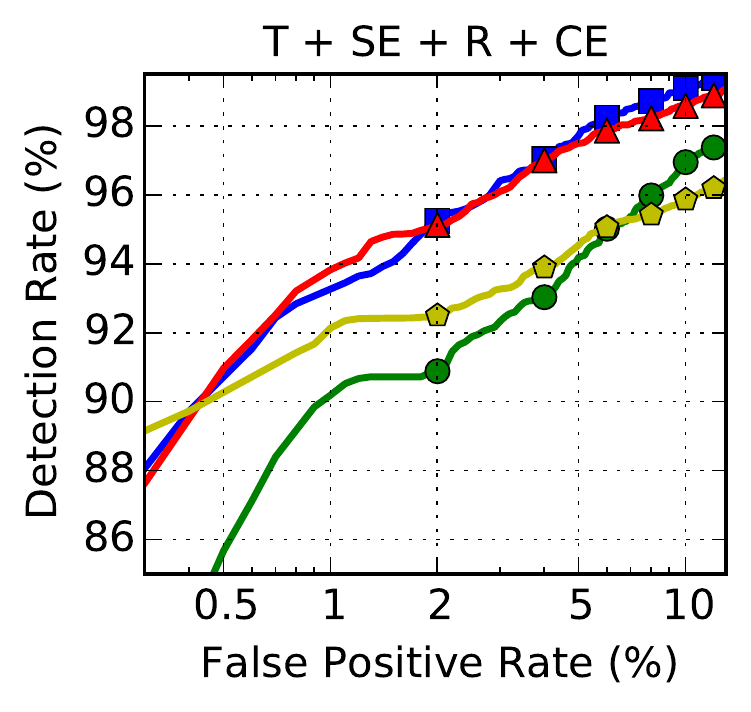}
\vspace{-5pt}
\caption{Mean ROC curves for all classifiers against different obfuscation techniques, computed on the \DContagio data.}\vspace{-5pt}
\label{fig:roc-drebin-obf}
\end{figure*}
\begin{figure*}[t]
\centering
\includegraphics[width=0.22\textwidth]{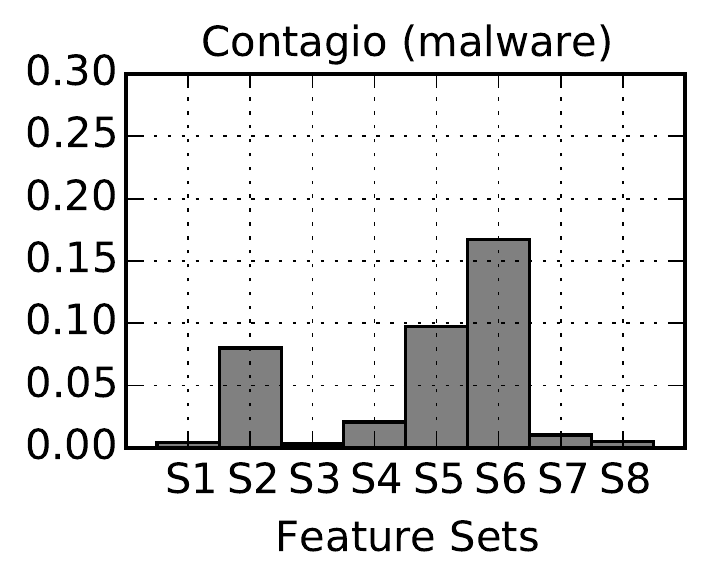}
\includegraphics[width=0.22\textwidth]{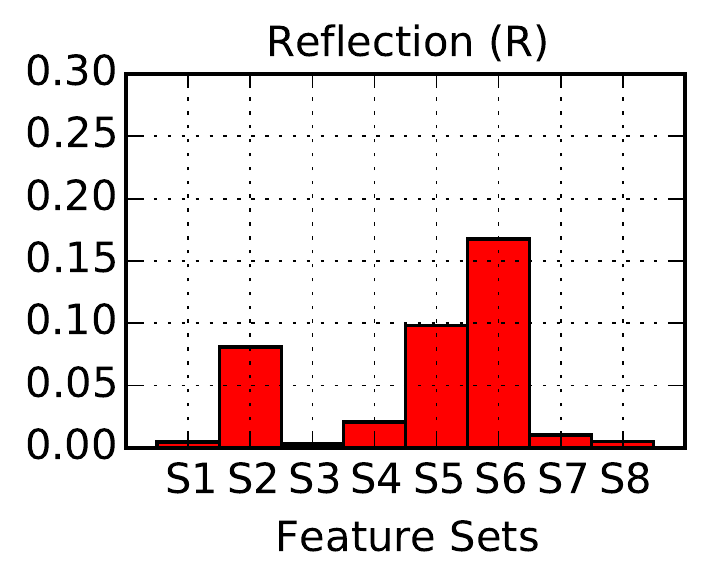}
\includegraphics[width=0.22\textwidth]{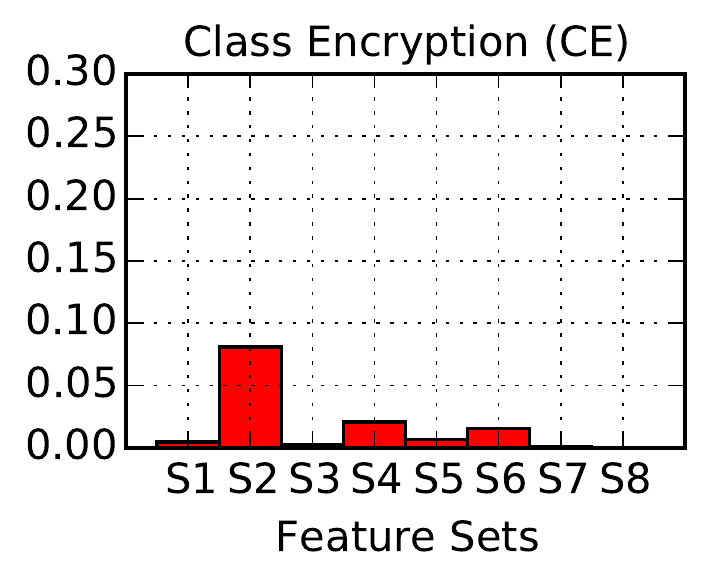}
\includegraphics[width=0.22\textwidth]{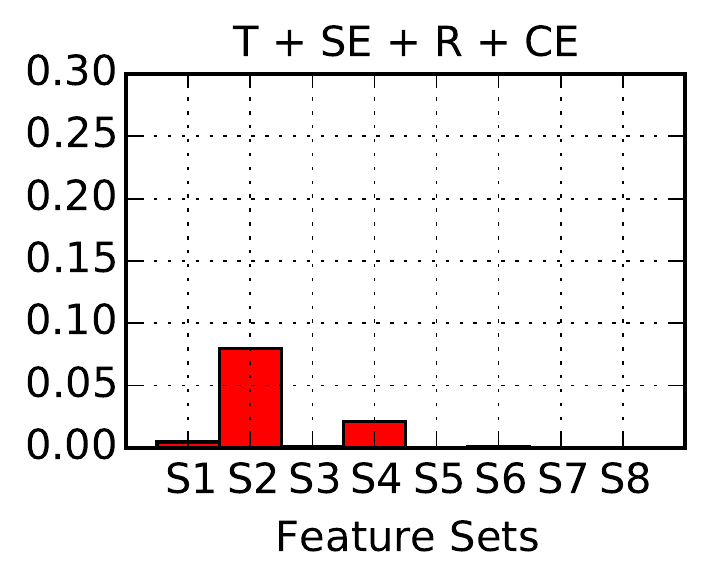}
\vspace{-5pt}
\caption{Fraction of features equal to one in each set (averaged on 10 runs), for non-obfuscated (\emph{leftmost} plot) and obfuscated malware in \DContagio, with different obfuscation techniques. While obfuscation deletes \dexcode features (S5-S8), the \manifest (S1-S4) remains mostly intact.}
\vspace{-8pt}
\label{fig:wstats-drebin-obfuscation}
\end{figure*}

\begin{figure*}[t]
\centering
\includegraphics[width=0.28\textwidth]{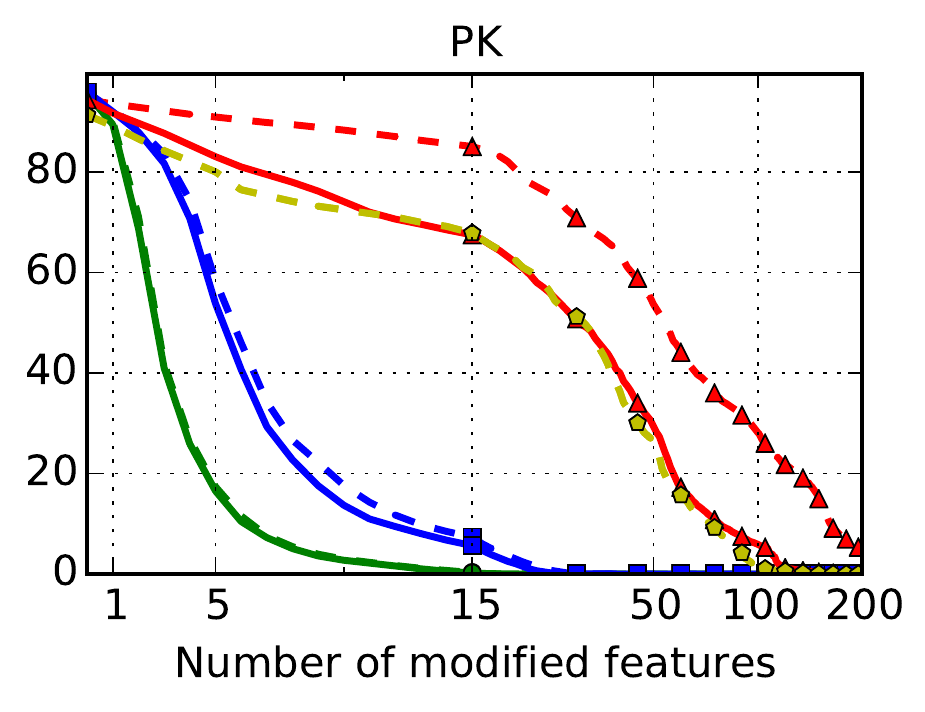}
\includegraphics[width=0.28\textwidth]{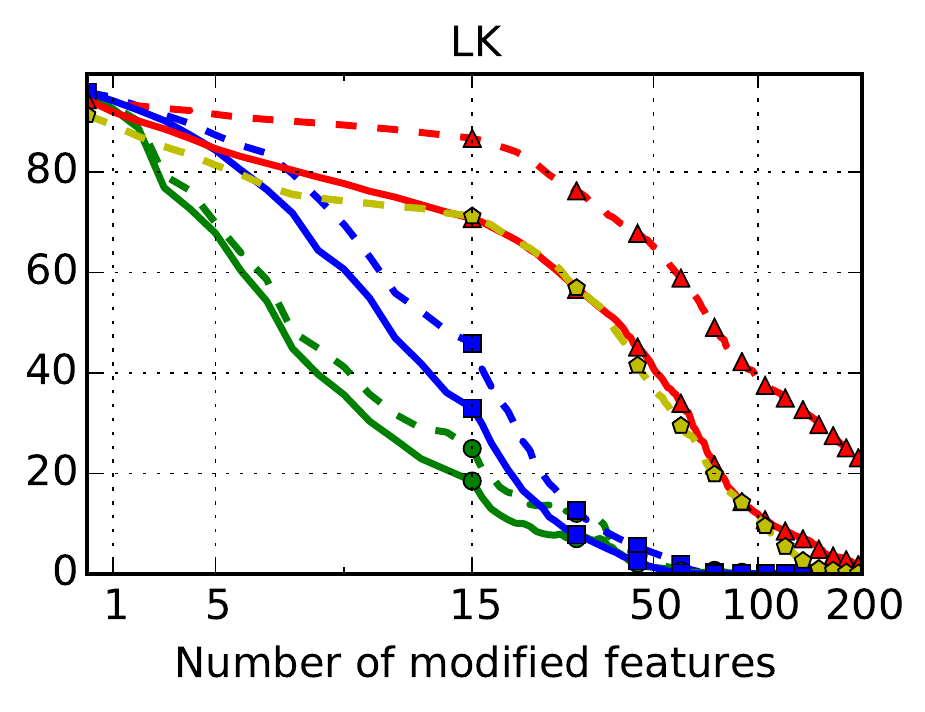}
\includegraphics[width=0.28\textwidth]{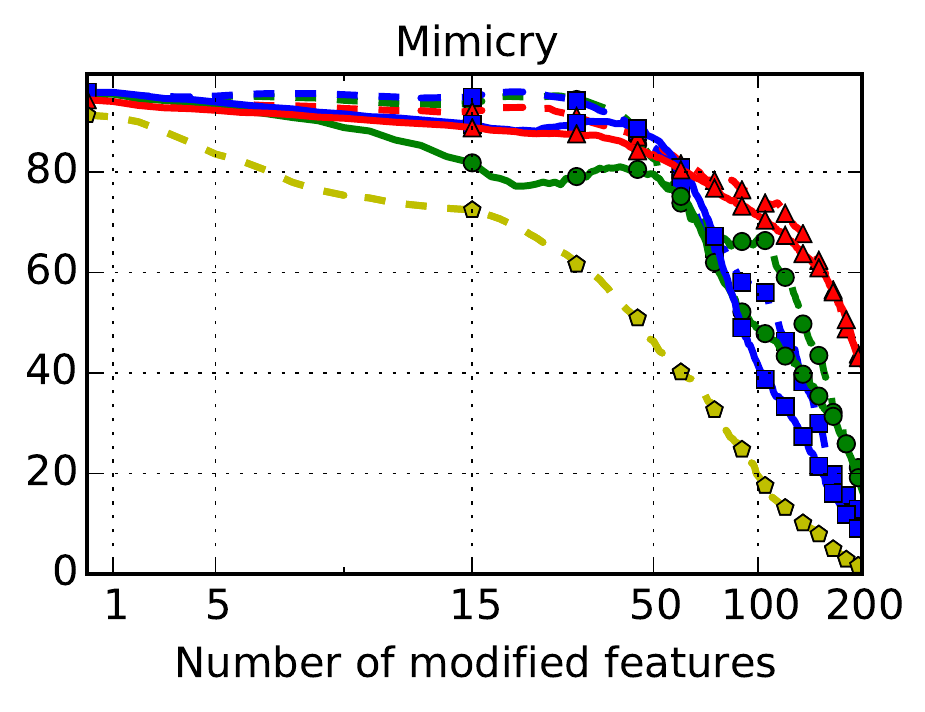}
\raisebox{15.5mm}{\includegraphics[width=0.13\textwidth]{figs/legend-vertical.pdf} }
\vspace{-5pt}
\caption{Detection Rate (DR) at 1\% False Positive Rate (FPR) for each classifier under the \emph{Perfect-Knowledge} (left), \emph{Limited-Knowledge} (middle), and \emph{Mimicry} (right) attack scenarios, against an increasing number of modified features. Solid (dashed) lines are obtained by simulating attacks with feature addition (feature addition and removal).}\vspace{-5pt}
\label{fig:worst_case_attack_results}
\end{figure*}

\begin{figure*}[t]
	\centering
	\includegraphics[width=0.22\textwidth]{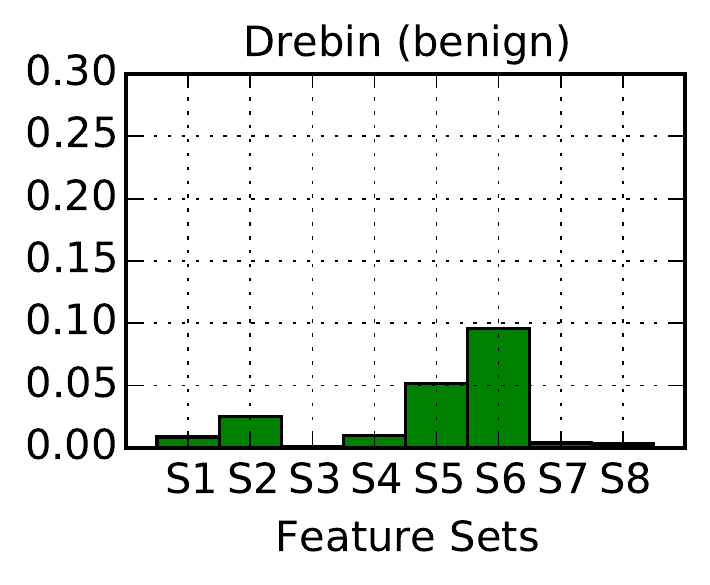}
	\includegraphics[width=0.22\textwidth]{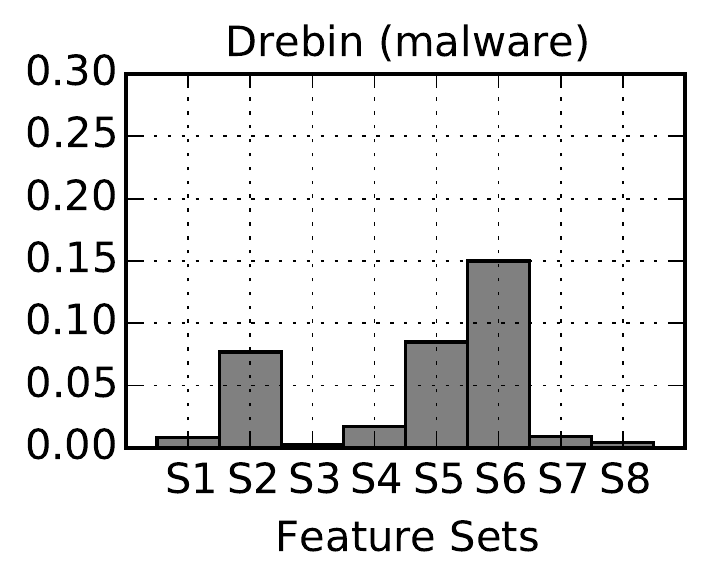}
	\includegraphics[width=0.22\textwidth]{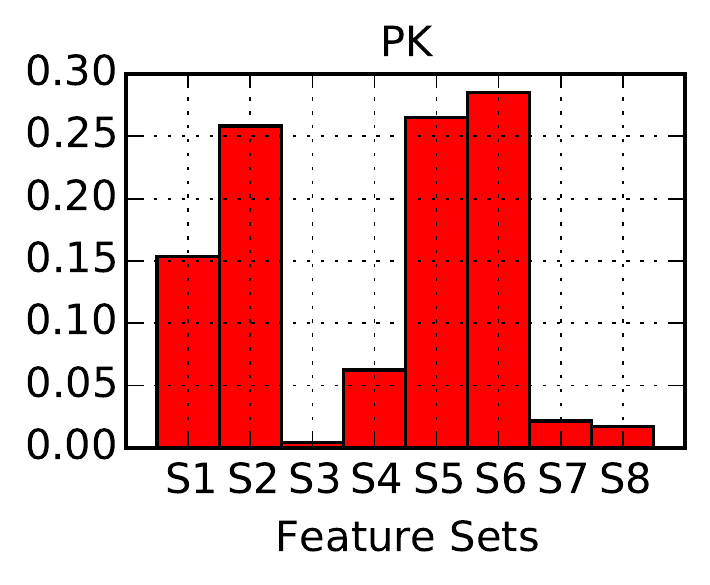}
	\includegraphics[width=0.22\textwidth]{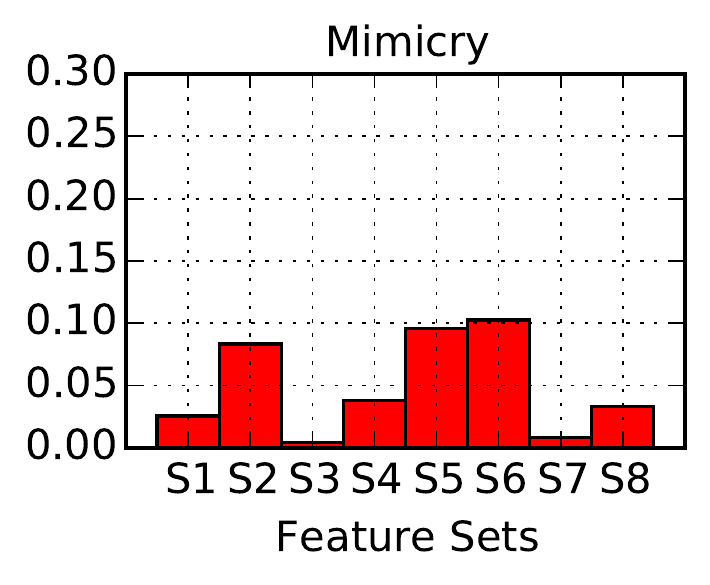}
	\vspace{-10pt}
	\caption{Fraction of features equal to one in each set (averaged on 10 runs) for benign (\emph{first} plot), non-obfuscated (\emph{second} plot) and DexGuard-based obfuscated malware in \DDrebin, using PK (\emph{third} plot) and mimicry (\emph{fourth} plot) attacks. It is clear that the mimicry attack produces malware samples which are more similar to the benign data than those obtained with the PK attack.}\vspace{-8pt}
	\label{fig:wstats-drebin-evasion}
\end{figure*}


\begin{table*}[t]
\centering
\caption{Top 5 modified features by the PK evasion attack with feature addition (A) and removal (R), for \SVM, \MCS, and \SSVM (highlighted in bold). The probability of a feature being equal to one in malware data is denoted with $p$.
For each classifier and each feature, we then report two values (averaged on 10 runs): ($i$) the probability  $q^{\prime}$ that the feature is modified by the attack (\emph{left}), and ($ii$) its relevance (\emph{right}), measured as its absolute weight divided by $\| \vct w \|_{1}$. If the feature is not modified within the first 200 changes, we report that the corresponding values are only lower than the minimum ones observed. 
In the last column, we also report whether the feature has been added ($\uparrow$) or removed ($\downarrow$) by the attack.}
\vspace{-8pt}
\label{tab:feat-mod}
\resizebox{\textwidth}{!}{%
\begin{tabular}{@{}clcllllllc@{}}
\toprule
\textbf{Feature Set} & \textbf{Feature Name} & $p$ & \multicolumn{2}{c}{\textbf{\SVM}} & \multicolumn{2}{c}{\textbf{\MCS}} & \multicolumn{2}{c}{\textbf{\SSVM}} & \textbf{A} ($\uparrow$) / \textbf{R} ($\downarrow$) \\ \midrule
S6                     & susp\_calls::android/telephony/gsm/SmsMessage;$\rightarrow$getDisplayMessageBody   & 2.40\%       & \textbf{89.60\%}          & 0.25\%          & 3.99\%           & 0.05\%          & $<$0.03\%          & $<$0.02\%         & $\uparrow$                                   \\
S1                     & req\_perm::android.permission.USE\_CREDENTIALS                                    & 0.05\%       & \textbf{65.77\%}          & 0.18\%          & \textbf{67.57\%}          & 0.13\%          & $<$0.03\%          & $<$0.02\%         & $\uparrow$                                   \\
S1                     & req\_perm::android.permission.WRITE\_OWNER\_DATA                                  & 0.52\%       & \textbf{64.76\%}          & 0.16\%          & \textbf{49.23\%}          & 0.11\%          & $<$0.03\%          & $<$0.02\%         & $\uparrow$                                   \\
S0                     & features::android.hardware.touchscreen                                                   & 0.60\%       & \textbf{64.01\%}          & 0.14\%          & 41.75\%          & 0.09\%          & $<$0.03\%          & $<$0.02\%         & $\uparrow$                                   \\
S6                     & susp\_calls::android/telephony/gsm/SmsMessage;$\rightarrow$getMessageBody          & 3.50\%       & \textbf{60.13\%}          & 0.13\%          & 17.30\%          & 0.05\%          & $<$0.03\%          & $<$0.02\%         & $\uparrow$                                   \\
S3                     & intent\_filters::android.intent.action.SENDTO                                            & 0.73\%       & 55.70\%          & 0.13\%          & \textbf{60.38\%}          & 0.11\%          & $<$0.03\%          & $<$0.02\%         & $\uparrow$                                   \\
S6                     & susp\_calls::android/telephony/CellLocation;$\rightarrow$requestLocationUpdate     & 0.05\%       & 50.87\%          & 0.12\%          & \textbf{48.37\%}          & 0.08\%          & $<$0.03\%          & $<$0.02\%         & $\uparrow$                                   \\
S6                     & susp\_calls::android/net/ConnectivityManager;$\rightarrow$getBackgroundDataSetting & 0.51\%       & 28.86\%          & 0.07\%          & \textbf{43.59\%}          & 0.09\%          & $<$0.03\%          & $<$0.02\%         & $\uparrow$                                   \\
S6                     & susp\_calls::android/telephony/TelephonyManager;$\rightarrow$getNetworkOperator    & 46.41\%      & 36.08\%          & 0.17\%          & 43.12\%          & 0.19\%          & \textbf{39.88\%}            & 0.04\%            & $\downarrow$                                 \\
S6                     & susp\_calls::android/net/NetworkInfo;$\rightarrow$getExtraInfo                     & 24.81\%      & 19.25\%          & 0.17\%          & 13.42\%          & 0.10\%          & \textbf{10.25\%}            & 0.03\%            & $\downarrow$                                 \\
S6                     & susp\_calls::getSystemService                                                      & 93.44\%      & $<$3.53\%        & $<$0.02\%       & $<$0.12\%        & $<$0.05\%       & \textbf{11.02\%}            & 0.02\%            & $\downarrow$                                 \\
S7                     & urls::www.searchmobileonline.com                                                         & 9.42\%       & 4.59\%           & 0.11\%          & 6.91\%           & 0.13\%          & \textbf{4.83\%}             & 0.03\%            & $\downarrow$                                 \\
S6                     & services::com.apperhand.device.android.AndroidSDKProvider                                & 10.83\%      & 6.78\%           & 0.13\%          & 4.19\%           & 0.09\%          & \textbf{5.14\%}             & 0.03\%            & $\downarrow$                                 \\
 \bottomrule
\end{tabular}%
}\vspace{-8pt}
\end{table*}

\myparagraph{DexGuard-based obfuscation attacks} The ROC curves reported in Fig.~\ref{fig:roc-drebin-obf} show the performance of the given classifiers, trained on \DDrebin, against the \texttt{DexGuard}-based obfuscation attacks (see Sect.~\ref{sect:obf} and Sect.~\ref{sect:obfuscation_techniques}) on the \DContagio malware. Here, \SSVM performs similarly to \MCS, while \SVM and \SSVM (M) typically exhibit lower detection rates.
Nevertheless, as these obfuscation attacks do not completely obfuscate the malware code, and the feature changes induced by them are not specifically targeted against any of the given classifiers, the classification performances are not significantly affected. In fact, the DR at $1\%$ FPR is never lower than $90\%$. As expected (see Sect.~\ref{sect:obfuscation_techniques}), strategies such as Trivial, String Encryption and Reflection do not affect the system performances significantly, as \Drebin only considers system-based API calls, which are not changed by the aforementioned obfuscations. Among these attacks, Class Encryption is the most effective strategy, as it is the only one that more significantly modifies the S5 and S7 feature sets (in particular, the first one), as it can be seen in Fig.~\ref{fig:wstats-drebin-obfuscation}. Nevertheless, even in this case, as \manifest-related features are not affected by \texttt{DexGuard}-based obfuscations, \Drebin still exhibits good detection performances.

\myparagraph{Advanced evasion} We finally report results for the PK, LK, and mimicry attacks in Fig.~\ref{fig:worst_case_attack_results}, considering both feature addition, and feature addition and removal.
As we are not removing \manifest-related features, \SSVM (M) is clearly tested only against feature-addition attacks.
Worth noting, \SSVM can drastically improve security compared to the other classifiers, as its performance decreases more gracefully against an increasing number of modified features, especially in the PK and LK attack scenarios.
In the PK case, while the DR of \Drebin (\SVM) drops to 60\% after modifying only \emph{two} features, the DR of the \SSVM decreases to the same amount only when \emph{fifteen} feature values are changed.
This means that our \SSVM approach can improve classifier security of about \emph{ten} times, in terms of the amount of modifications required to create a malware sample that evades detection.
The underlying reason is that \SSVM provides more evenly-distributed feature weights, as shown in Fig.~\ref{fig:wdist-drebin}. Note that \SSVM and \SSVM (M) exhibit a maximum absolute weight value of 0.5 (on average). This means that, in the worst case, modifying a single feature yields an average decrease of the classification function equal to 0.5, while for \MCS and \SVM this decrease is approximately 1 and 2.5, respectively. It is thus clear that, to achieve a comparable decrease of the classification function (\ie, a comparable probability of evading detection), more features should be modified in the former cases.
Finally, it is also worth noting that mimicry attacks are less effective, as expected, as they exploit an inferior level of knowledge of the targeted system. Despite this, an interesting insight on the behavior of such attacks is reported in Fig.~\ref{fig:wstats-drebin-evasion}. After modifying a large number of features, the mimicry attack tends to produce a distribution that is very close to that of the benign data (even without removing any \manifest-related feature). This means that, in terms of their feature vectors, benign and malware samples become very similar. Under these circumstances, no machine-learning technique can separate benign and malware data with satisfying accuracy. The vulnerability of the system may be thus regarded as intrinsic in the choice of the feature representation, rather than in how the classification function is learned. This clearly confirms the importance of designing features that are more difficult to manipulate for an attacker.

\myparagraph{Feature manipulation} To provide some additional insights, in Table~\ref{tab:feat-mod} we report the top 5 modified features by the PK attack with feature addition and removal for \SVM, \MCS, and \SSVM. For each classifier, we select the top 5 features by ranking them in descending order of the probability of modification $q^{\prime}$. This value is computed as follows.
First, the probability $q$ of modifying the $k^{\rm th}$ feature in a malware sample, regardless of the maximum number of admissible modifications, is computed as:
\begin{equation}
q = \mathbb{E}_{ \vct x  \sim p( \vct x | y=+1)} \{  x_{k} \neq x^{\prime}_{k} \} = p^{\nu} (1-p)^{1-\nu}  \, ,
\end{equation}
where $\mathbb{E}$ denotes the expectation operator, $p( \vct x | y=+1)$ the distribution of malware samples,
$x_{k}$ and $x^{\prime}_{k}$ are the $k^{\rm th}$ feature values before and after manipulation,
and $p$ is the probability of observing $x_{k}=1$ in malware. Note that $\nu=1$ if $x_{k}=1$, $x_{k}$ does not belong to the \texttt{manifest} sets S1-S4, and the associated weight $\hat w_{k}>0$, while $\nu = 0$ if  $\hat w_{k}<0$ (otherwise the probability of modification $q$ is zero).
This formula denotes compactly that, if a feature can be modified, then it will be changed with probability $p$ (in the case of deletion) or $1-p$ (in the case of insertion).
Then, to consider that features associated to the highest absolute weight values are modified more frequently by the attack, with respect to an increasing maximum number $m$ of modifiable features, we compute $q^{\prime} = \mathbb{E}_{m} \{q\}$. Considering $m=1, \ldots, \con d$, with uniform probability,
each feature will be modified with probability $q^{\prime}=q  \, (\con d - r)/ \con d $, with $r=0$ for the feature $x_{(1)}$ assigned to the highest absolute weight value, $r=1$ for the second ranked feature $x_{(2)}$, \etc 
In general, for the $k^{\rm th}$-ranked feature $x_{(k)}$, $r=k-1$, for $k=1, \ldots, \con d$.
Thus, $q^{\prime}$ decreases depending on the feature ranking, which in turn depends on the feature weights and the probability $p$ of the feature being present in malware.

Regarding Table~\ref{tab:feat-mod}, note first how the probability of modifying the top features, along with their \emph{relevance} (\ie, their absolute weight value with respect to $\| \vct w \|_{1}$), decreases from \SVM to \MCS, and from \MCS to \SSVM. 
These two observations are clearly connected. The fact that the attack modifies features with a lower probability depends  on the fact that weights are more evenly distributed. To better understand this phenomenon, imagine the limit case in which all features are assigned the same absolute weight value. It is clear that, in this case, the attacker could randomly modify any subset of features and obtain the same effect on the classification output; thus, on average, each feature will have the same probability of being modified.

The probability of modifying a feature, however, does not only depend on the weight assigned by the classifier, but also on the probability of being present in malware data, as mentioned before. For instance, if a (non-manifest) feature is present in all malware samples, and it has been assigned a very high positive weight, it will be always removed; conversely, if it only rarely occurs in malware, then it will be deleted only from few samples.
This behavior is clearly exhibited by the top features modified by \SSVM. In fact, since this classifier basically assigns the same absolute weight value to almost all features, the top modified ones are simply those appearing more frequently in malware. More precisely, in our experiments this classifier, as a result of our parameter optimization procedure, assigns a higher (absolute) weight to features present in malware, and a lower (absolute) weight to features present in benign data (\ie, $| w_{k}^{\rm ub} | >  | w_{k}^{\rm lb} |$, $k=1,\ldots,\con d$).
This is why, conversely to \SVM and \MCS, the attack against \SSVM tends to remove features, rather then injecting them. To conclude, it is nevertheless worth pointing out that, in general, the most frequently-modified features clearly depend on the data distribution (\ie, on class imbalance, feature correlations, \etc), and not only on the probability of being more frequent in malware. In our analysis, this dependency is intrinsically captured by the dependency of $q^{\prime}$ on the feature weights learned by the classifier.

\myparagraph{Robustness and regularization} Interestingly, a recent theoretical explanation behind the fact that more features should be manipulated to evade our \SSVM can also be found in~\cite{xu09}. In particular, in that work Xu~\etal have shown that the \emph{regularized} SVM learning problem, as given in Eq.~\eqref{eq:obj-ssvm}, is equivalent to a non-regularized, \emph{robust} optimization problem, in which the input data is corrupted by a worst-case $\ell_{2}$ (spherical) noise. Note that this noise is \emph{dense}, as it tends to slightly affect all feature values. More generally, Xu~\etal~\cite{xu09} have shown that the regularization term depends on the kind of hypothesized noise over the input data. Our evasion attacks are \emph{sparse}, as the attacker aims to minimize the number of modified features, and thus they significantly affect only the most discriminant ones. This amounts to consider an $\ell_{1}$ worst-case noise over the input data. In this case, Xu \etal~\cite{xu09} have shown that the optimal regularizer is the $\ell_{\infty}$ norm of $\vct w$. In our \SSVM, the key idea is to add a box constraint on $\vct w$, as given in Eq.~\eqref{eq:constr-ssvm}, which is essentially equivalent to consider an additional $\ell_{\infty}$ regularizer on $\vct w$, consistently with the findings in \cite{xu09}.

\vspace{-5pt}
\section{Limitations and Open Issues} \label{sect:limitations}

Despite the very promising results achieved by our \SSVM, it is clear that such an approach exhibits some intrinsic limitations.
First, as \Drebin performs a static code analysis, it is clear that also \SSVM can be defeated by more sophisticated encryption and obfuscation attacks. However, it is also worth remarking that this is not a vulnerability of the learning algorithm itself, but rather of the chosen feature representation, and for this reason we have not considered these attacks in our work.
A similar behavior is observed when a large number of features is modified by our evasion attacks, and especially in the case of mimicry  attacks (see Sect. \ref{sect:exp}), in which the manipulated malware samples almost exactly replicate benign data (in terms of their feature vectors). This is again possible due to an intrinsic vulnerability of the feature representation, and no learning algorithm can clearly separate such data with satisfying accuracy.
Nevertheless, this problem only occurs when malware samples are significantly modified and, as pointed out in Sect.~\ref{sect:malware-manipulation}, it might be very difficult for the attacker to do that without compromising their intrusive functionality, or without leaving significant traces of adversarial manipulation.
For example, the introduction of changes such as reflective calls requires a careful manipulation of the Dalvik registers (\eg, verifying that old ones are correctly re-used and that new ones can be safely employed). A single mistake in the process can lead to verification errors, and the application might not be usable anymore (we refer the reader to \cite{hoffmann16-codaspy,hoffmann16-tr} for further details). 
Another limitation of our approach may be its unsatisfying performance under PK and LK attacks, but this can be clearly mitigated with simple countermeasures to prevent that the attacker gains sufficient knowledge of the attacked system, such as frequent system re-training and diversification of training data collection~\cite{biggio14-ijprai}.
To summarize, although our approach is clearly not bulletproof, we believe that it significantly improves the security of the baseline \Drebin system (and of the standard SVM algorithm).

\vspace{-8pt}
\section{Conclusions and Future Work}
\label{sect:conclusion}

Recent results in the field of adversarial machine learning and computer security have confirmed the intuition pointed out by Barreno~\etal~\cite{barreno06-asiaccs,barreno10,huang11}, namely, that machine learning itself can introduce specific vulnerabilities in a security system, potentially compromising the overall system security.
The underlying reason is that machine-learning techniques have not been originally designed to deal with intelligent and adaptive attackers, who can modify their behavior to mislead the learning and classification algorithms.

The goal of this work has been, instead, to show that machine learning \emph{can} be used to improve system security, if one follows an \emph{adversary-aware} approach that proactively anticipates the attacker.
To this end, we have first exploited a general framework for assessing the security of learning-based malware detectors, by modeling attackers with different goals, knowledge of the system, and capabilities of manipulating the data.
We have then considered a specific case study involving \Drebin, an Android malware detection tool, and shown that the performance of \Drebin can be significantly downgraded in the presence of skilled attackers that can carefully manipulate malware samples to evade classifier detection.
The main contribution of this work has been to define a novel, theoretically-sound learning algorithm to train linear classifiers with more evenly-distributed feature weights. This approach allows one to improve system security (in terms of requiring a much higher number of careful manipulations to the malware samples), without significantly affecting computational efficiency.

A future development of our work, which may further improve classifier security, is to extend our approach for secure learning to nonlinear classifiers, \eg, using \emph{nonlinear kernel functions}. Although \emph{nonlinear kernels} can not be directly used in our approach (due to the presence of a linear constraint on $\vct w$), one may exploit a trick known as the \emph{empirical kernel mapping}. It consists of first mapping samples onto an \emph{explicit} (approximate) kernel space, and then learning a linear classifier on that space~\cite{scholkopf99}. We would like to remark here that also investigating the trade-off between \emph{sparsity} and \emph{security} highlighted in Sect.~\ref{sect:secsvm-alg} may provide interesting insights for future work. In this respect, the recent findings in~\cite{xu09} related to robustness and regularization of learning algorithms (briefly summarized at the end of Sect.~\ref{sect:exp}) may provide inspiring research directions.

Another interesting future extension of our approach may be to explicitly consider, for each feature, a different level of robustness against the corresponding adversarial manipulations. In practice, however, the \emph{agnostic} choice of assuming equal robustness for all features may be preferred, as it may be very difficult to identify features that are more difficult to manipulate. If categorizing features according to their robustness to adversarial manipulations is deemed feasible, instead, then this knowledge may be incorporated into the learning algorithm, such that higher (absolute) weight values are assigned to more robust features. 

It is finally worth remarking that we have also recently exploited the proposed learning algorithm to improve the security of PDF and Javascript malware detection systems against \emph{sparse} evasion attacks~\cite{russu16-aisec,demontis16-spr}. This witnesses that our proposal does not only provide a first, concrete example of how machine learning can be exploited to improve \emph{security} of Android malware detectors, but also of how our design methodology can be readily applied to other learning-based malware detection tasks.

\vspace{-8pt}


\bibliographystyle{IEEEtranS}

\begin{IEEEbiography}
[{\includegraphics[width=1in,height =1.25in,clip,keepaspectratio]{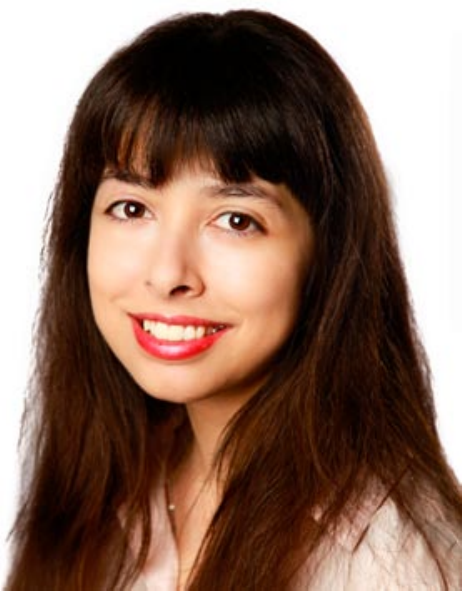}}]{Ambra Demontis (S'16)} received the M.Sc. degree in Information Technology with honors from the University of Cagliari, Italy, in 2014. She is now a Ph.D. student in Department of Electrical and Electronic Engineering, University of Cagliari. Her current  research interests include machine learning, computer security and biometrics.
She is a student member of the IEEE and of the IAPR.
\end{IEEEbiography} \vspace{-15pt}

\begin{IEEEbiography}
[{\includegraphics[width=1in,height =1.25in,clip,keepaspectratio]{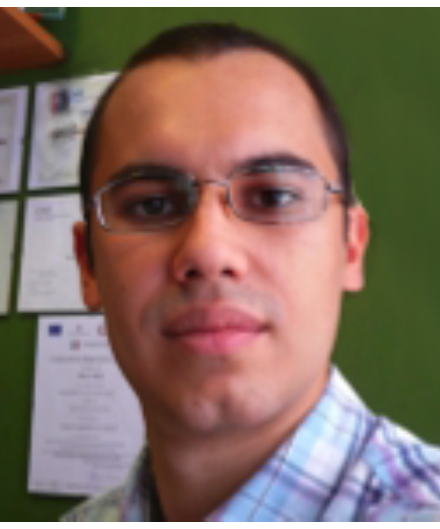}}]{Marco Melis (S'16)} received his B.Sc. degree in Electronic Engineering from the University of Cagliari, Italy, in February 2014. Since 2014, he has been with the Department of Electrical and Electronic Engineering, University of Cagliari. His research activity is focused on machine learning, kernel methods and biometric recognition systems. Marco Melis is a student member of the IEEE.
\end{IEEEbiography} \vspace{-10pt}

\begin{IEEEbiography}
[{\includegraphics[width=1in,height =1.25in,clip,keepaspectratio]{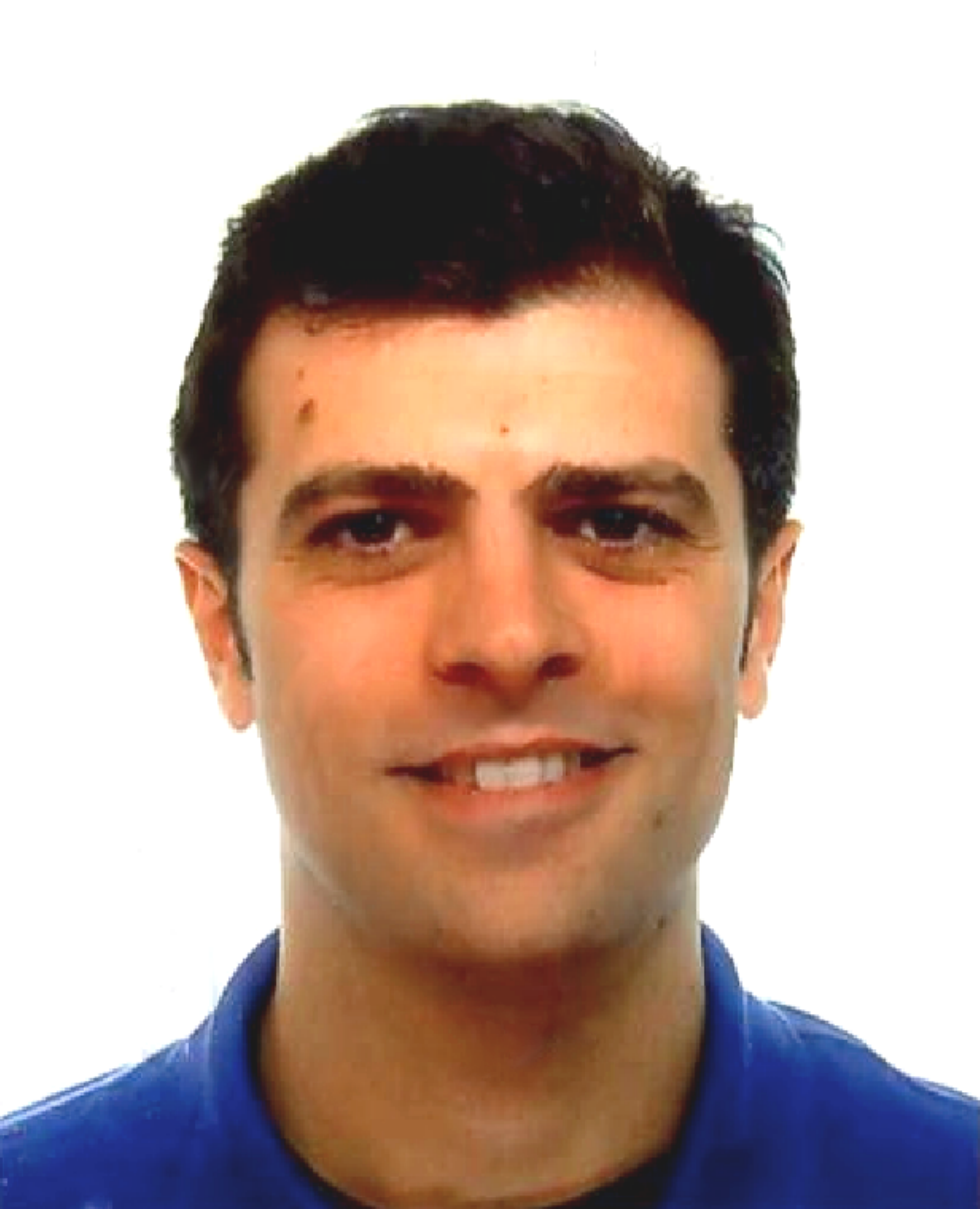}}]{Battista Biggio (SM'17)} received the M.Sc. degree (Hons.) in Electronic Engineering and the Ph.D. degree in Electronic Engineering and Computer Science from the University of Cagliari, Italy, in 2006 and 2010. Since 2007, he has been with the Department of Electrical and Electronic Engineering, University of Cagliari, where he is currently an Assistant Professor. In 2011, he visited the University of T\"ubingen, Germany, and worked on the security of machine learning to training data poisoning. His research interests include secure machine learning, multiple classifier systems, kernel methods, biometrics and computer security. Dr. Biggio serves as a reviewer for several international conferences and journals. He is a senior member of the IEEE and member of the IAPR.
\end{IEEEbiography} \vspace{-8pt}

\begin{IEEEbiography}
[{\includegraphics[width=1in,height =1.25in,clip,keepaspectratio]{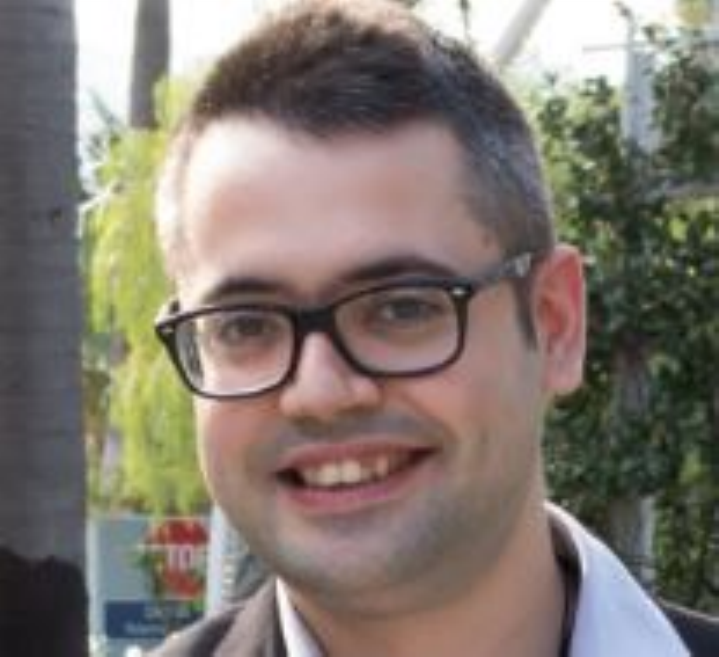}}]{Davide Maiorca (M'16)} received the M.Sc. degree (Hons.) in Electronic Engineering and the Ph.D. in Electronic and Computer Engineering from the University of Cagliari, Italy, respectively in 2012 and 2016. In 2013, he visited the Systems Security group at Ruhr-Universit\"at Bochum, guided by Prof. Dr. Thorsten Holz, and worked on advanced obfuscation of Android malware. He now holds a Post-Doctoral position at the University of Cagliari. His current research interests include adversarial machine learning, malware in documents and Flash applications, Android malware and mobile fingerprinting. He has been a member of the 2016 IEEE Security \& Privacy Student Program Committee.
\end{IEEEbiography}

\begin{IEEEbiography}
[{\includegraphics[width=1in,height =1.25in,clip,keepaspectratio]{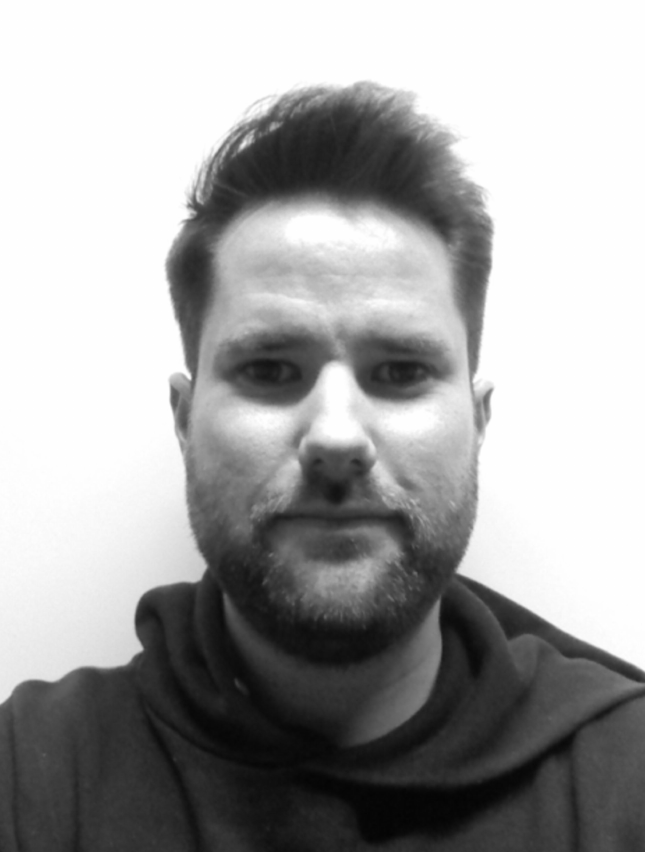}}]{Daniel Arp} is a PhD candidate working in the Computer Security Group at
Technische Universit{\"a}t Braunschweig. He received his diploma degree
in Computer Engineering from the Technische Universit{\"a}t Berlin in
2012 and worked as a research assistant at the University
of G{\"o}ttingen subsequently. His research focuses on the application of
learning methods for malware detection and privacy protection.
\end{IEEEbiography} \vspace{-8pt}

\begin{IEEEbiography}
[{\includegraphics[width=1in,height =1.25in,clip,keepaspectratio]{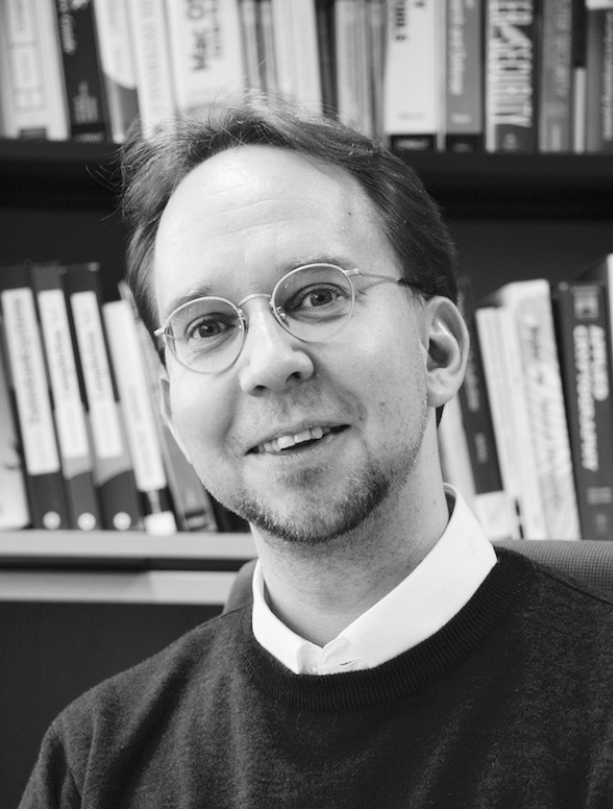}}]{Konrad Rieck} is a Professor at Technische Universit{\"a}t
Braunschweig, where he leads the Institute of System Security.
Prior to taking this position, he has been working at the
University of G{\"o}ttingen, Technische Universit{\"a}t Berlin
and Fraunhofer Institute FIRST. He graduated from Freie
Universit{\"a}t Berlin in 2004 and received a Doctorate in
Computer Science from Technische Universit{\"a}t Berlin in 2009.
Konrad Rieck is a recipient of the CAST/GI Dissertation Award
IT-Security and a Google Faculty Research Award. His interests
revolve around computer security and machine learning, including
the detection of computer attacks, the analysis of malicious
code, and the discovery of vulnerabilities.
\end{IEEEbiography} \vspace{-8pt}

\begin{IEEEbiography}
[{\includegraphics[width=1in,height =1.25in,clip,keepaspectratio]{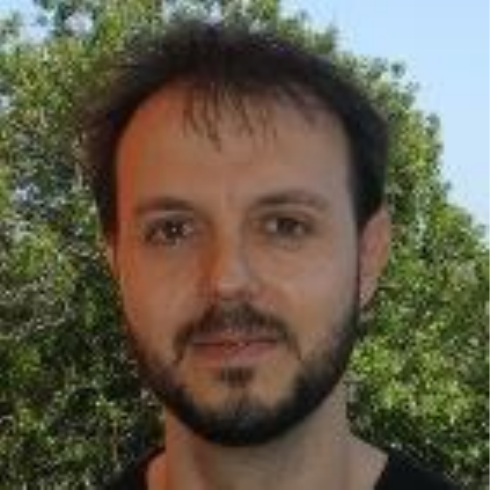}}]{Igino Corona} received the M.Sc. in Electronic Engineering and Ph.D. in Computer Engineering at the University of Cagliari, Italy, in 2006 and 2010, respectively. In 2009 he also worked at the Georgia Tech Information Security Center as a research scholar. He is currently a post-doctoral researcher at the University of Cagliari. His main research interests include web security, detection of fast flux networks, adversarial machine learning and web intrusion detection.
\end{IEEEbiography} \vspace{-8pt}

\begin{IEEEbiography}[{\includegraphics[width=1in,height =1.25in,clip,keepaspectratio]{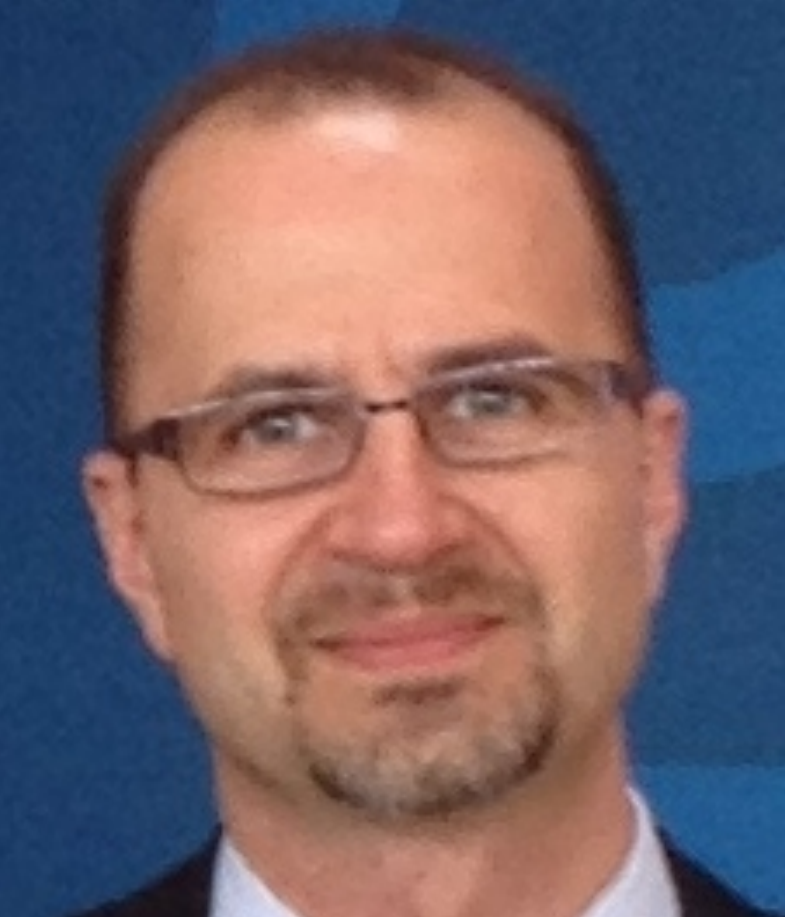}}]{Giorgio Giacinto (SM'10)}  is Associate Professor of Computer Engineering at the University of Cagliari, Italy. He obtained the MS degree in Electrical Engineering in 1994, and the Ph.D. degree in Computer Engineering in 1999. Since 1995 he joined the research group on Pattern Recognition and Applications of the DIEE, University of Cagliari, Italy. His research interests are in the area of pattern recognition and its application to computer security, and image classification and retrieval. During his career Giorgio Giacinto has published more than 120 papers on international journals, conferences, and books. He is a senior member of the ACM and the IEEE. He has been involved in the scientific coordination of several research projects in the fields of pattern recognition and computer security, at the local, national and international level. Since 2012, he has been the organizer of the Summer School on Computer Security and Privacy ``Building Trust in the Information Age'' (\url{http://comsec.diee.unica.it/summer-school}).
\end{IEEEbiography}

\begin{IEEEbiography}[{\includegraphics[width=1in,height =1.25in,clip,keepaspectratio]{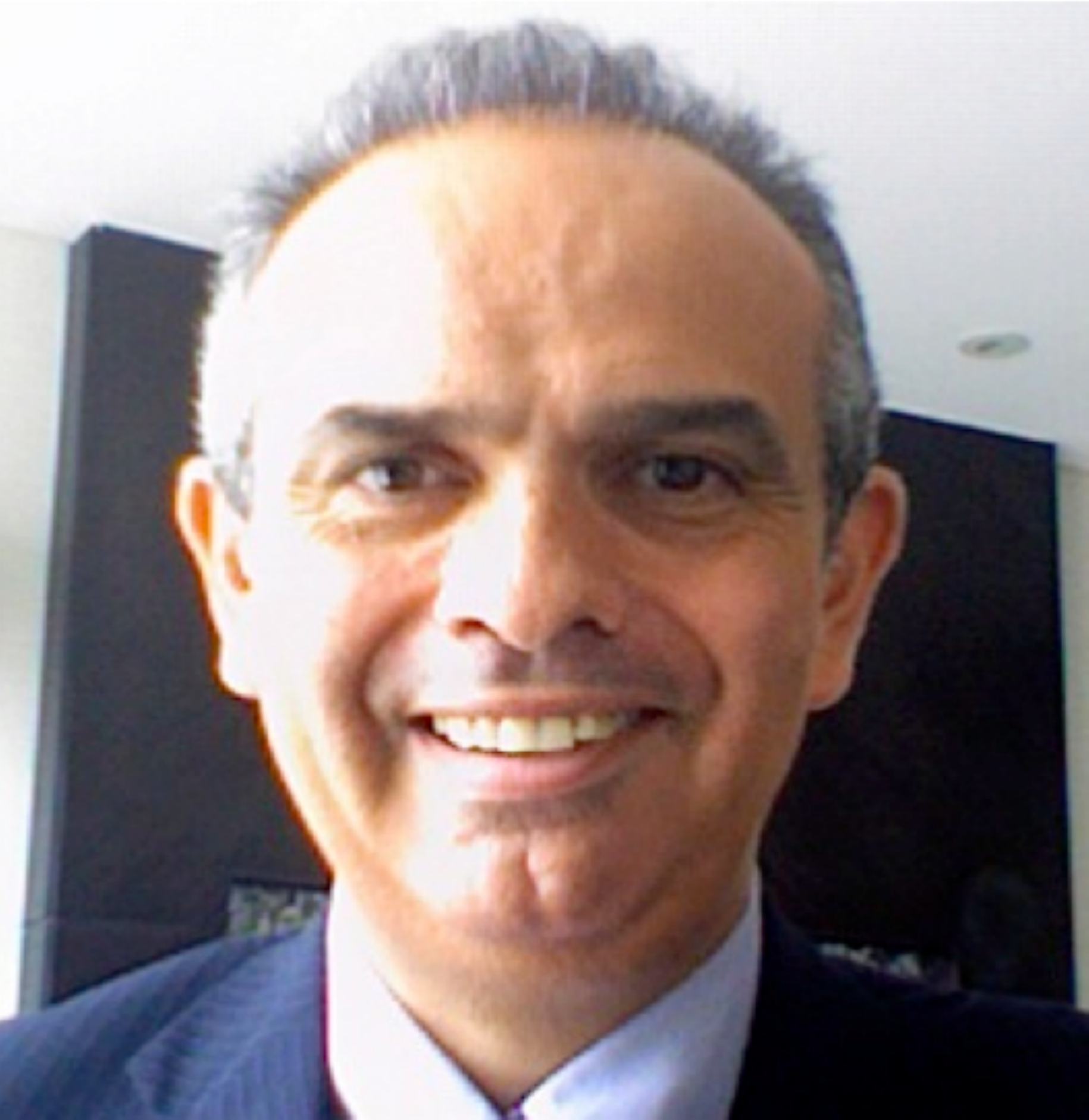}}]{Fabio Roli (F'12)}  received his Ph.D. in Electronic Engineering from the University of Genoa, Italy. He was a research group member of the University of Genoa (88-94). He was adjunct professor at the University of Trento ('93-'94). In 1995, he joined the Department of Electrical and Electronic Engineering of the University of Cagliari, where he is now professor of Computer Engineering and head of the research laboratory on pattern recognition and applications. His research activity is focused on the design of pattern recognition systems and their applications. He was a very active organizer of international conferences and workshops, and established the popular workshop series on multiple classifier systems. Dr. Roli is Fellow of the IEEE and of the IAPR.
\end{IEEEbiography}

\end{document}